\RequirePackage{fix-cm}
\documentclass[smallextended,final,nospthms]{svjour3}       
\smartqed  
\usepackage{graphicx}
\usepackage{mathptmx}      
\DeclareMathAlphabet{\mathcal}{OMS}{cmsy}{m}{n}
\usepackage{amsmath}
\usepackage{amsthm}
\usepackage{amssymb}
\usepackage[ruled, linesnumbered, noend]{algorithm2e}
\usepackage{tikz}
\theoremstyle{plain}
\newtheorem{mytheorem}{Theorem}
\newtheorem{mylemma}{Lemma}
\newtheorem{mycorollary}{Corollary}
\newtheorem{myproposition}{Proposition}

\theoremstyle{definition}
\newtheorem{mydefinition}{Definition}

\newcommand{\pcl}{\Omega}
\newcommand{\MS}{\Delta}
\newcommand{\PC}{\Pi}
\newcommand{\ecc}{\texttt{cc}}
\newcommand{\vc}{\texttt{\textbf{vc}}}

\newcommand{\mw}{\texttt{\textbf{mw}}}
\newcommand{\CC}{\mathcal{C}}
\newcommand{\w}{\mathcal{W}}
\newcommand{\MC}{\text{MC}}
\newcommand{\Tr}{\text{Tr}}
\journalname{Algorithmica}
\begin{document}

\title{Finding Optimal Triangulations Parameterized by Edge Clique Cover\thanks{This work has been financially supported by Academy of Finland (grant 322869).}
}


\author{Tuukka Korhonen}


\institute{Department of Computer Science, University of Helsinki, Helsinki, Finland\\
              \email{tuukka.korhonen@uib.no}           
}

\date{ }

\maketitle

\begin{abstract}
We consider problems that can be formulated as a task of finding an optimal triangulation of a graph w.r.t. some notion of optimality.
We present algorithms parameterized by the size of a minimum edge clique cover ($\ecc$) to such problems.
This parameterization occurs naturally in many problems in this setting, e.g., in the perfect phylogeny problem $\ecc$ is at most the number of taxa, in fractional hypertreewidth $\ecc$ is at most the number of hyperedges, and in treewidth of Bayesian networks $\ecc$ is at most the number of non-root nodes.

We show that the number of minimal separators of graphs is at most $2^\ecc$, the number of potential maximal cliques is at most $3^\ecc$, and these objects can be listed in times $O^*(2^\ecc)$ and $O^*(3^\ecc)$, respectively, even when no edge clique cover is given as input; the $O^*(\cdot)$ notation omits factors polynomial in the input size.
These enumeration algorithms imply $O^*(3^\ecc)$ time algorithms for problems such as treewidth, weighted minimum fill-in, and feedback vertex set.
For generalized and fractional hypertreewidth we give $O^*(4^m)$ time and $O^*(3^m)$ time algorithms, respectively, where $m$ is the number of hyperedges.
When an edge clique cover of size $\ecc'$ is given as a part of the input we give $O^*(2^{\ecc'})$ time algorithms for treewidth, minimum fill-in, and chordal sandwich.
This implies an $O^*(2^n)$ time algorithm for perfect phylogeny, where $n$ is the number of taxa.
We also give polynomial space algorithms with time complexities $O^*(9^{\ecc'})$ and $O^*(9^{\ecc + O(\log^2 \ecc)})$ for problems in this framework.
\keywords{Parameterized algorithms \and Potential maximal cliques \and Edge clique cover \and Treewidth \and Minimum fill-in \and Fractional hypertreewidth}
\end{abstract}

\section{Introduction}
A graph is \emph{chordal} if it has no induced cycle of length at least four.
A \emph{triangulation} of a graph $G$ is a chordal supergraph of $G$ on the same vertex set.
Many graph problems can be formulated as a problem of finding an optimal triangulation of the graph with respect to some notion of optimality.
For example, computing the treewidth of a graph corresponds to finding a triangulation with the smallest size of a maximum clique, and computing the minimum fill-in corresponds to finding a triangulation with the least number of edges.

An \emph{edge clique cover} of a graph $G$ is a set of cliques of $G$ such that each edge of $G$ is contained in at least one of the cliques.
In this article, we give fixed-parameter algorithms for optimal triangulation problems parameterized by the size of a minimum edge clique cover of the graph, denoted by $\ecc$, and by the size of an edge clique cover given as an input, denoted by $\ecc'$.
Our algorithms are based on the framework of potential maximal cliques~\cite{DBLP:journals/siamcomp/BouchitteT01,DBLP:journals/siamcomp/FominKTV08,DBLP:journals/siamcomp/FominTV15,DBLP:conf/lata/Gysel14,DBLP:journals/ipl/MollTT12}.

A \emph{minimal triangulation} of a graph $G$ is a triangulation of $G$ that has no subgraph that is a triangulation of $G$.
A \emph{potential maximal clique} (PMC) of a graph $G$ is a set of vertices $\pcl \subseteq V(G)$ such that there exists a minimal triangulation $H$ of $G$ where $\pcl$ is a maximal clique.
By the results of Bouchitt{\'{e}} and Todinca~\cite{DBLP:journals/siamcomp/BouchitteT01} and Fomin, Kratsch, Todinca, and Villanger~\cite{DBLP:journals/siamcomp/FominKTV08}, a large class of optimal triangulation problems can be solved by first enumerating all PMCs of the graph and then performing dynamic programming over them, with time complexity that is linear in the number of PMCs and polynomial in the size of the graph.
Therefore the main focus of this article is on bounding the number of PMCs based on edge clique cover and giving a corresponding enumeration algorithm.

\subsection{Interpretations of $\ecc$}
While in general the parameter $\ecc$ could be considered non-standard, it has natural interpretations in at least three settings in which algorithms for finding optimal triangulations are applied: width parameters of hypergraphs, phylogenetics, and probabilistic inference.
The reason that $\ecc$ is a natural choice in these settings is that the input graph is constructed as a union of cliques, with the goal of each clique $W$ representing a constraint of type ``the triangulation must contain a maximal clique $\pcl$ with $W \subseteq \pcl$''.
Maximal cliques of a triangulation in turn correspond to bags of a tree decomposition.
Next we discuss these three settings in more detail.
The discussion is summarized in Table~\ref{table:param_rela}.

\begin{table}[!t]
\begin{center}
\begin{tabular}{|p{6.2cm}|p{3.3cm}|p{0.9cm}|}
\hline
Problem & Parameter & Relation \\
\hline
Computing width parameters of hypergraphs: primal treewidth, generalized hypertreewidth, and fractional hypertreewidth & Number of hyperedges $m$ & $\ecc \le m$\\
\hline
Perfect phylogeny problem and its optimization variant & Number of taxa $n$ & $\ecc \le n$\\
\hline
Computing treewidth of a moral graph of Bayesian network & Number of non-root nodes $n'$ & $\ecc \le n'$\\
\hline
\end{tabular}
\end{center}
\caption{Relationships between the parameter minimum edge clique cover ($\ecc$) and natural parameters in optimal triangulation problems.}
\label{table:param_rela}
\end{table}

A \emph{hypergraph} is a structure like a graph, but instead of edges that are sets of two vertices, a hypergraph has hyperedges that are arbitrary subsets of the vertex set.
Width parameters of hypergraphs, including primal treewidth, generalized hypertreewidth, and fractional hypertreewidth are central structural parameters of constraint satisfaction problems (CSPs)~\cite{DBLP:journals/jacm/GottlobMS09,DBLP:journals/talg/GroheM14}.
In particular, they measure how well a hypergraph can be decomposed by cuts whose intersection with a solution has a simple characterization, allowing dynamic programming.
The computation of each of these parameters can be formulated as an optimal triangulation problem on the \emph{primal graph} of the hypergraph~\cite{DBLP:journals/ipl/MollTT12}.
The primal graph of a hypergraph is a graph constructed by inducing a clique on each hyperedge, and therefore the size of its minimum edge clique cover is at most $m$, the number of hyperedges.
For example, the primal treewidth of a hypergraph is the treewidth of the primal graph.
Generalized and fractional hypertreewidth are more general parameters, but their definitions are technical and postponed to Section~\ref{subsec:defs}.

In phylogenetics a central problem is to construct an evolutionary tree of a set of $n$ taxa (i.e. species) based on $k$ characters (i.e. attributes) describing them~\cite{semple2003phylogenetics}.
For example, when the character data is drawn from molecular sequences, the number of characters $k$ can be much larger than the number of taxa $n$~\cite{DBLP:journals/siamcomp/KannanW97}.
In the weighted minimum fill-in problem the input is a graph $G$ and a weight function on pairs of vertices of $G$, and the task is to find a triangulation $H$ of $G$ so that the total weight of edges in $E(H) \setminus E(G)$ is minimized.
Deciding if the taxa admit a perfect phylogeny and an optimization version of it can be reduced to the weighted minimum fill-in problem on the partition intersection graph of the characters~\cite{DBLP:journals/dm/BordewichHS05,DBLP:conf/lata/Gysel14,semple2003phylogenetics}.
The partition intersection graph has a vertex for each character-state pair, and its edges are constructed by inducing a clique corresponding to each taxon~\cite{semple2003phylogenetics}.
Hence the size of its minimum edge clique cover is at most $n$, the number of taxa.

A third setting in which parameterization by $\ecc$ is motivated is probabilistic inference.
Given a Bayesian network, the first step of efficient probabilistic inference algorithms is to compute a tree decomposition of small width of the moral graph of the Bayesian network~\cite{DBLP:conf/uai/JensenJ94,nielsen2009bayesian}.
The moral graph is constructed as a union of $n'$ cliques, where $n'$ is the number of non-root nodes of the Bayesian network~\cite{nielsen2009bayesian}, and thus the size of its minimum edge clique cover is at most $n'$.

\subsection{Connections to Practice}
\label{sec:practice}
\begin{figure}[!b]
\centering
\includegraphics[width=0.5\linewidth]{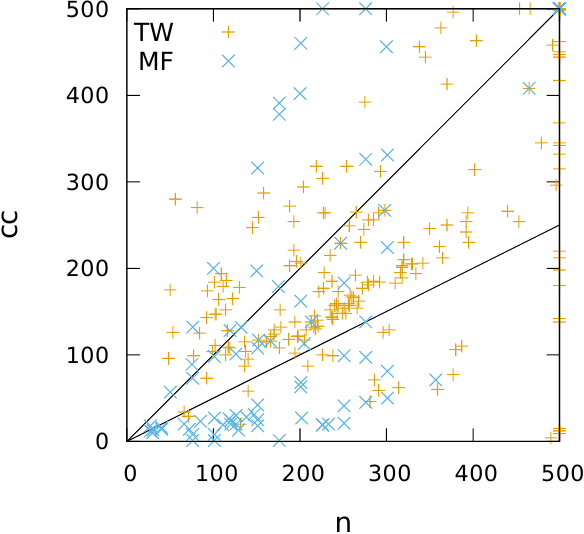}
\caption{Number of vertices ($n$) and minimum edge clique cover ($\ecc$) of PACE 2017 treewidth (TW) and minimum fill-in (MF) instances. Both values are truncated from above at 500 and the lines $n = \ecc$ and $n/2 = \ecc$ are shown.}
\label{fig:ecc_vs_n}
\end{figure}
This article is also directly motivated by observations in practice.
Starting from the Second Parameterized Algorithms and Computational Experiments challenge (PACE 2017)~\cite{DBLP:conf/iwpec/DellKTW17}, algorithm implementations based on potential maximal cliques have been observed to outperform other exact algorithm implementations on problems formulated as finding optimal triangulations~\cite{korhonen2020finding,DBLP:journals/jea/KorhonenBJ19,DBLP:conf/aaai/KorhonenJ20,DBLP:conf/pods/RavidMK19,DBLP:conf/sea2/Tamaki19,DBLP:journals/jco/Tamaki19}.
In particular, this work is motivated by experimental observations of the usefulness of potential maximal cliques in computing hypergraph parameters~\cite{korhonen2020finding,DBLP:journals/jea/KorhonenBJ19} and in phylogenetics~\cite{korhonen2020finding,DBLP:conf/aaai/KorhonenJ20}.

Our parameterization can be justified by real-world instances with small edge clique covers.
In the context of hypergraph parameters, 708 of the 3072 hypergraphs in the standard HyperBench library~\cite{DBLP:conf/pods/FischlGLP19} have $m < n/2$, where $n$ is the number of vertices and $m$ is the number of hyperedges.
In the context of phylogenetics, an instance describing mammal mitochondrial sequences~\cite{hasegawa1985dating} has 7 taxa while its partition intersection graph has 245 vertices and an instance describing Indo-European languages~\cite{ringe2002indo} has 24 taxa while its partition intersection graph has 864 vertices.
In the context of Bayesian networks, the Bayesian network ``Andes''~\cite{conati1997line}, accessed from the standard BNlearn repository~\cite{bnlearn}, has 223 nodes of which 134 are non-root.

To study the exact values of $\ecc$ on well-known treewidth and minimum fill-in benchmarks, we computed minimum edge clique covers of graphs from PACE 2017~\cite{DBLP:conf/iwpec/DellKTW17}, which had tracks for both treewidth and minimum fill-in.
The treewidth track had 200 available test instances and the minimum fill-in track 100.
All of the instances are based on real-world applications~\cite{DBLP:conf/iwpec/DellKTW17}.
We attempted to compute minimum edge clique covers of all of the 300 instances using Bron--Kerbosch algorithm~\cite{DBLP:journals/cacm/BronK73} for maximal clique enumeration and CPLEX~\cite{cplex} for solving the resulting set cover problem.
We managed to compute the minimum edge clique cover of 294 of the instances.
Figure~\ref{fig:ecc_vs_n} shows the relations of minimum edge clique cover and the number of vertices in these instances.
Most of the instances have minimum edge clique cover between $n/2$ and $n$, where $n$ is the number of vertices.
In 75 of the instances $\ecc < n/2$, which is roughly the asymptotic threshold where our bounds for minimal separators and potential maximal cliques are better than known bounds by $n$~\cite{DBLP:journals/siamcomp/FominTV15}.

\subsection{Techniques}
The algorithms that we design in this article are based on the framework of potential maximal cliques (PMCs)~\cite{DBLP:journals/siamcomp/BouchitteT01,DBLP:journals/siamcomp/FominKTV08}.
Algorithms in this framework typically consist of two phases.
In the first phase the set $\PC(G)$ of PMCs of the input graph $G$ is enumerated, and in the second phase dynamic programming over the PMCs is performed in time $O^*(|\PC(G)|)$.
The second phase of the PMC framework has already been formulated for all of the problems that we consider~\cite{DBLP:journals/siamcomp/FominKTV08,DBLP:journals/siamcomp/FominTV15,DBLP:journals/tcs/FuruseY14,DBLP:conf/lata/Gysel14,DBLP:journals/ipl/MollTT12}, so our $O^*(3^\ecc)$ time algorithms follow from an $O^*(3^\ecc)$ time PMC enumeration algorithm that we give.
This algorithm is based on the Bouchitt{\'{e}}--Todinca algorithm~\cite{DBLP:journals/tcs/BouchitteT02}.
We achieve the $O^*(3^\ecc)$ bound by novel characterizations of minimal separators and PMCs with respect to an edge clique cover.
In particular, we show that minimal separators correspond to bipartitions of an edge clique cover and almost all potential maximal cliques correspond to tripartitions of an edge clique cover.

On some of the problems we improve the time complexity to $O^*(2^{\ecc'})$, where $\ecc'$ is the size of an edge clique cover given as an input.
The $O^*(2^{\ecc'})$ time algorithms use the same dynamic programming states as the standard PMC framework, but instead of using PMCs for transitions we use fast subset convolution~\cite{DBLP:conf/stoc/BjorklundHKK07}.
The application of fast subset convolution requires ad-hoc techniques for each problem to take into account the cost caused by the PMC implicitly selected by the convolution.

We also give algorithms that work in polynomial space and in times $O^*(9^{\ecc'})$ and $O^*(9^{\ecc+O(\log^2 \ecc)})$.
These algorithms are based on a polynomial space and $O^*(9^\ecc)$ time algorithm for enumerating PMCs and on a lemma asserting that every minimal triangulation of a graph $G$ has a maximal clique that is in a sense a balanced separator with respect to an edge clique cover of $G$.

\subsection{Contributions}
\label{sec:contributions}
\paragraph{Combinatorial bounds and enumeration algorithms.}
We start by giving bounds for the numbers of minimal separators and PMCs.
We use $S(n, k)$ to denote the Stirling numbers of the second kind, i.e., the number of ways to partition an $n$-element set into $k$ non-empty subsets.

\begin{mytheorem}
\label{the:numbers}
If $G$ is a graph with an edge clique cover of size $\ecc$, then the number of minimal separators of $G$ is at most $S(\ecc, 2)$ and the number of potential maximal cliques of $G$ is at most $S(\ecc, 3) + \ecc$.
\end{mytheorem}
We also show that these bounds are exactly tight for all values of $\ecc$.
Note that $S(\ecc, 2) \le 2^\ecc$ and $S(\ecc, 3) + \ecc \le 3^\ecc$.

We use $\MS(G)$ to denote the set of minimal separators of a graph $G$.
There are $O^*(|\MS(G)|)$ time algorithms for enumerating the minimal separators of $G$~\cite{DBLP:journals/ijfcs/BerryBC00,DBLP:journals/dam/Takata10}, so it follows that the minimal separators of a graph can be enumerated in $O^*(2^\ecc)$ time.
For enumerating the PMCs $\PC(G)$ of a graph $G$ no algorithms that are linear in the size of the output $|\PC(G)|$ and polynomial in the size of the input are known\footnote{Obtaining an output-linear input-polynomial time algorithm for enumerating PMCs has been explicitly stated as an open problem in 2006~\cite{bodlaender2006open} and to the best of our knowledge is still open.}.
Despite that, we are able to design an efficient algorithm for enumerating PMCs parameterized by $\ecc$, even when no edge clique cover is given as input.
The fact that the algorithm works even when no edge clique cover is given as input is crucial because there is no $O^*(2^{2^{o(\ecc)}})$ time parameterized algorithm for minimum edge clique cover assuming the exponential time hypothesis~\cite{DBLP:journals/siamcomp/CyganPP16}.

\begin{mytheorem}
\label{the:pmc_enum}
There is an algorithm that given a graph $G$ whose minimum edge clique cover has size $\ecc$ enumerates the potential maximal cliques of $G$ in $O^*(3^\ecc)$ time.
\end{mytheorem}

\paragraph{Corollaries of Theorem~\ref{the:pmc_enum}.}
From Theorem~\ref{the:pmc_enum} it follows that all problems that admit $O^*(|\PC(G)|)$ time algorithms when the set $\PC(G)$ of PMCs of the input graph $G$ is given can be solved in $O^*(3^\ecc)$ time, even when no edge clique cover is given as an input.
Next we briefly discuss these corollaries, summarized in Table~\ref{tab:cor}.

\begin{table}[!t]
\begin{center}
\begin{tabular}{|p{5cm}|p{2.8cm}|p{1.9cm}|}
\hline
Problem & Parameter & Time complexity \\
\hline
Computing treewidth~\cite{DBLP:journals/siamcomp/FominKTV08} & Minimum edge clique cover $\ecc$ & $O^*(3^\ecc)$\\
\hline
Computing (weighted) minimum fill-in~\cite{DBLP:journals/siamcomp/FominKTV08} & Minimum edge clique cover $\ecc$ & $O^*(3^\ecc)$\\
\hline
Finding an induced subgraph with treewidth $\le t$ satisfying a CMSO formula $\phi$~\cite{DBLP:journals/siamcomp/FominTV15} & Minimum edge clique cover $\ecc$ & $O^*(3^\ecc \cdot f(t, \phi))$\\
\hline
Computing fractional hypertreewidth of a hypergraph~\cite{DBLP:journals/ipl/MollTT12} & Number of hyperedges $m$ & $O^*(3^m)$\\
\hline
\end{tabular}
\end{center}
\caption{Parameterized algorithms obtained as corollaries of Theorem~\ref{the:pmc_enum}.\label{tab:cor}}
\end{table}

Computing treewidth and minimum fill-in are classical triangulation problems~\cite{DBLP:journals/siamcomp/BouchitteT01}.
The treewidth of a graph is the minimum possible maximum clique size in a triangulation of the graph, minus one.
The minimum fill-in of a graph is the minimum number of edges to add to make a graph chordal.
In weighted minimum fill-in, the input, in addition to the graph, includes a weight function assigning non-negative weights to the potential edges to add, and the task is to minimize the sum of the weights of the edges added.
Our $O^*(3^\ecc)$ time algorithms for treewidth and (weighted) minimum fill-in follow from Theorem~\ref{the:pmc_enum} and the dynamic programming of Fomin et al.~\cite{DBLP:journals/siamcomp/FominKTV08}.

Theorem~\ref{the:pmc_enum} implies corollaries for all problems in a framework called \emph{maximum induced subgraph of bounded treewidth}, including for example the problems maximum independent set, minimum feedback vertex set, and longest induced path~\cite{DBLP:journals/siamcomp/FominTV15}.
In particular, using the dynamic programming of~\cite{DBLP:journals/siamcomp/FominTV15} as a black box, we obtain $O^*(3^\ecc \cdot f(t, \phi))$ time algorithms for problems that can be expressed in a following form, where $t$ is an integer and $\phi$ is a counting monadic second order logic formula: Given a graph $G$, find a maximum size vertex subset $X$ so that there exists a vertex subset $F$ with $X \subseteq F \subseteq V(G)$, the treewidth of the induced subgraph $G[F]$ is at most $t$, and it holds that $(G[F], X) \models \phi$.

Computing the fractional hypertreewidth of a hypergraph corresponds to finding a triangulation of its primal graph, minimizing the maximum fractional edge cover of a maximal clique in the triangulation (see Section~\ref{subsec:defs} for a complete definition).
By combining the fact that a primal graph of a hypergraph has edge clique cover of size $m$, the number of hyperedges, with Theorem~\ref{the:pmc_enum} and the dynamic programming algorithm of~\cite{DBLP:journals/ipl/MollTT12}, we obtain an $O^*(3^m)$ time algorithm for fractional hypertreewidth.

\paragraph{Optimizing the results in special cases.}
For some triangulation problems we obtain better results than would directly follow from Theorem~\ref{the:pmc_enum}.

A straightforward application of Theorem~\ref{the:pmc_enum} would result in an $O^*(6^m)$ time algorithm for generalized hypertreewidth (see definition in Section~\ref{subsec:defs}).
We optimize this a bit.
\begin{mytheorem}
\label{the:ghtw}
Given a hypergraph with $m$ hyperedges, its generalized hypertreewidth can be computed in $O^*(4^m)$ time.
\end{mytheorem}

When an edge clique cover of size $\ecc'$ is given as an input, some of the algorithms can be optimized to $O^*(2^{\ecc'})$ time.
Chordal sandwich is a special case of the weighted minimum fill-in problem, where the task is to decide if there exists a fill-in with weight zero.
We pay attention to it because it has application to the perfect phylogeny problem~\cite{DBLP:journals/dm/BordewichHS05,DBLP:conf/lata/Gysel14,semple2003phylogenetics}.

\begin{mytheorem}
\label{the:fast}
Given a graph with an edge clique cover of size $\ecc'$, its treewidth and minimum fill-in can be computed in $O^*(2^{\ecc'})$ time, and also the chordal sandwich problem on it can be solved in $O^*(2^{\ecc'})$ time.
\end{mytheorem}

\begin{mycorollary}[See \cite{DBLP:conf/lata/Gysel14}]
The perfect phylogeny problem can be solved in $O^*(2^n)$ time, where $n$ is the number of taxa.
\end{mycorollary}

A previous parameterized algorithm for perfect phylogeny works in time $O^*(4^r)$, where $r \le n$ is the arity of characters~\cite{DBLP:journals/siamcomp/KannanW97}.
Our $O^*(2^n)$ time algorithm improves over it in the case when $r > n/2$.
This case is motivated by the fact that the $O^*(4^r)$ algorithm works for partial characters only via a reduction that sets $r \ge nf$ for a fraction $f$ of missing data~\cite{semple2003phylogenetics}.

\paragraph{Polynomial space algorithms.}
We also give polynomial space algorithms for some of the problems.
The algorithms work in $O^*(9^{\ecc'})$ time when an edge clique cover of size $\ecc'$ is given as an input and in $O^*(9^{\ecc + O(\log^2 \ecc)})$ time when the parameter $\ecc$ is given as an input.

\begin{mytheorem}
\label{the:poly_alg1}
There is a polynomial space $O^*(9^{\ecc'})$ time algorithm for treewidth and weighted minimum fill-in, where $\ecc'$ is the size of an edge clique cover given as an input.
There are also polynomial space $O^*(9^m)$ time algorithms for both generalized and fractional hypertreewidth, where $m$ is the number of hyperedges.
\end{mytheorem}

Our polynomial space result for the parameter $\ecc$ is weaker than Theorem~\ref{the:pmc_enum} in the sense that it requires the integer $\ecc$ as an input.

\begin{mytheorem}
\label{the:poly_alg2}
There is a polynomial space $O^*(9^{\ecc + O(\log^2 \ecc)})$ time algorithm for treewidth and minimum fill-in, where $\ecc$ is an integer given as an input that is at least the size of a minimum edge clique cover of the input graph.
\end{mytheorem}

The present article extends the preliminary work~\cite{korhonen:LIPIcs:2020:13325} in three ways.
First, Theorem~\ref{the:numbers} has been improved so that we have lower and upper bounds that match exactly on all values of $\ecc$.
Second, Theorem~\ref{the:ghtw} which provides a non-trivial extension of our results to generalized hypertreewidth has been added.
Third, data on minimum edge clique covers of PACE 2017 instances has been added to Section~\ref{sec:practice}.

\subsection{Related Work}
\label{sec:relwork}
The prior fixed-parameter (FPT) algorithms for PMC enumeration include an $O^*(4^\vc)$ time algorithm, where $\vc$ is the size of a minimum vertex cover, and an $O^*(1.7347^\mw)$ time algorithm, where $\mw$ is the modular width~\cite{DBLP:journals/algorithmica/FominLMT18}, extending the $O(1.7347^n)$ time algorithm, where $n$ is the number of vertices~\cite{DBLP:journals/siamcomp/FominTV15}.
One can see that edge clique cover and vertex cover are orthogonal parameters by considering complete graphs and star graphs.
For modular width we show in Section~\ref{sec:mw} that the relation $\mw \le 2^\ecc-2$ holds, but there are graphs with $\mw = 2^\ecc-2$.
In the conclusion of~\cite{DBLP:journals/algorithmica/FominLMT18} the authors mentioned that they are not aware of other FPT parameters than $\vc$ and $\mw$ for PMCs and asked whether more parameterizations could be obtained.

Other parameterized approaches on the PMC framework include an FPT modulator parameter~\cite{DBLP:journals/algorithmica/LiedloffMT19} and a slicewise polynomial (XP) bound for minimal separators in $H$-graphs~\cite{DBLP:conf/esa/FominGR18}.
The modulator parameter is orthogonal to edge clique cover.
On $H$-graphs, the graphs with edge clique cover of size $\ecc$ are $K_{\ecc}$-graphs, so the $H$-graph parameterization implies an $n^{O(\ecc^2)}$ time algorithm for enumerating PMCs.

In addition to the $O^*(4^r)$ time algorithm for perfect phylogeny~\cite{DBLP:journals/siamcomp/KannanW97} that we discussed in Section~\ref{sec:contributions}, we are not aware of prior single-exponential FPT algorithms with the same parameters as our algorithms.
For fractional hypertreewidth there are parameterized algorithms whose parameters depend on the sizes of intersections of hyperedges~\cite{DBLP:conf/pods/FischlGP18}.
For treewidth and chordal sandwich, different techniques have been used to obtain an $O^*(3^\vc)$ time algorithm for treewidth~\cite{DBLP:journals/dam/ChapelleLTV17} and an $O^*(2^{\vc'})$ time algorithm for chordal sandwich, where $\vc'$ is the size of a minimum vertex cover of the admissible edge set~\cite{DBLP:conf/ciac/HeggernesMNV10}.

\subsection{Organization of the Article}
In Section~\ref{sec:preli} we give necessary definitions and background on minimal triangulations and PMCs.
In Section~\ref{sec:ccco} we characterize minimal separators and PMCs based on edge clique cover, proving Theorem~\ref{the:numbers}.
In Section~\ref{sec:enum_alg} we give enumeration algorithms for PMCs, proving Theorem~\ref{the:pmc_enum}.
In Section~\ref{sec:ghtw} we give the algorithm for generalized hypertreewidth, proving Theorem~\ref{the:ghtw}.
In Section~\ref{sec:fast} we give faster algorithms for the case when an edge clique cover is given as an input, proving Theorem~\ref{the:fast}.
In Section~\ref{sec:poly} we give polynomial space algorithms, proving Theorems~\ref{the:poly_alg1} and~\ref{the:poly_alg2}.
In Section~\ref{sec:tight} we show that our combinatorial bounds are tight.
In Section~\ref{sec:mw} we prove the relation of edge clique cover and modular width claimed in Section~\ref{sec:relwork}.
We conclude in Section~\ref{sec:conclusion}.

\section{Preliminaries}
\label{sec:preli}
We recall the standard graph notation that we use and preliminaries on minimal triangulations.
We also give formal definitions of the problems that we consider and introduce our notation related to edge clique cover.

\subsection{Notation on Graphs}
We consider graphs that are finite, simple, and undirected.
We assume that the graphs given as input are connected.
For graphs with multiple connected components, the algorithms can be applied to each connected component independently.
The sets of vertices and edges of a graph $G$ are denoted by $V(G)$ and $E(G)$, respectively.
The set of edges of a complete graph with vertex set $X$ is denoted as $X^2$.
The subgraph $G[X]$ induced by $X \subseteq V(G)$ has $V(G[X]) = X$ and $E(G[X]) = E(G) \cap X^2$.
We also use the notation $G \setminus X = G[V(G) \setminus X]$.
The vertex sets of connected components of a graph $G$ are $\CC(G)$.
The set of neighbors of a vertex $v$ is denoted by $N(v)$ and the set of neighbors of a vertex set $X$ by $N(X) = \bigcup_{v \in X} N(v) \setminus X$.
The closed neighborhood of a vertex $v$ is $N[v] = N(v) \cup \{v\}$ and the closed neighborhood of a vertex set $X$ is $N[X] = N(X) \cup X$.
A clique of a graph $G$ is a vertex set $X$ such that $G[X]$ is complete.
The set of inclusion maximal cliques of $G$ is denoted by $\MC(G)$.

\subsection{Minimal Triangulations}
A graph is \emph{chordal} if it has no induced cycle of four or more vertices.
A chordal graph $H$ is a \emph{triangulation} of a graph $G$ if $V(G) = V(H)$ and $E(G) \subseteq E(H)$.
A triangulation $H$ of $G$ is a \emph{minimal triangulation} of $G$ if there is no other triangulation $H'$ of $G$ with $E(H') \subsetneq E(H)$.
The edges in $E(H) \setminus E(G)$ are \emph{fill-edges}.
A vertex set $\pcl \subseteq V(G)$ is a \emph{potential maximal clique} (PMC) of $G$ if there is a minimal triangulation $H$ of $G$ such that $\pcl \in \MC(H)$.
The set of PMCs of $G$ is denoted by $\PC(G)$.

A vertex set $S$ is a \emph{minimal $a,b$-separator} of graph $G$ if the vertices $a$ and $b$ are in different components of $G \setminus S$, and $S$ is inclusion minimal in this regard.
A \emph{full component} of a vertex set $X$ is a component $C \in \CC(G \setminus X)$ with $N(C) = X$.
We note that $S$ is a minimal $a,b$-separator if and only if $S$ has distinct full components containing $a$ and $b$.
A vertex set $S$ is a \emph{minimal separator} if it is a minimal $a,b$-separator for some pair $a,b$, i.e., it has at least two full components.
We denote the set of minimal separators of $G$ with $\MS(G)$.

A \emph{block} of a graph $G$ is a vertex set $C \subseteq V(G)$ such that $G[C]$ is connected and $N(C) \in \MS(G)$, i.e., $C$ is a full component of some minimal separator.
We remark that a common notation is to call a pair $(N(C), C)$ a full block~\cite{DBLP:journals/siamcomp/BouchitteT01}.
In modern formulations of the PMC framework the concept of non-full blocks is not needed~\cite{DBLP:journals/siamcomp/FominTV15}, so we simplify the notation by identifying the block with only the vertex set $C$.

Next we recall a couple of required propositions on the structure of PMCs.
Figure~\ref{fig:pmc_ex} provides examples giving intuition about the propositions.

\begin{myproposition}[\cite{DBLP:journals/siamcomp/BouchitteT01}]
\label{pro:pmc_basic}
A vertex set $\pcl \subseteq V(G)$ is a PMC of a graph $G$ if and only if
\begin{enumerate}
\item $N(C) \subsetneq \pcl$ for all $C \in \CC(G \setminus \pcl)$, i.e., no component of $\pcl$ is full, and
\item for all pairs of distinct vertices $u,v \in \pcl$, either $\{u, v\} \in E(G)$ or there is a component $C \in \CC(G \setminus \pcl)$ with $\{u, v\} \subseteq N(C)$.
\end{enumerate}
\end{myproposition}

We will refer to condition~1 of Proposition~\ref{pro:pmc_basic} as the \emph{no full component condition} and to condition~2 as the \emph{cliquish condition}.
Note that Proposition~\ref{pro:pmc_basic} implies an $O(nm)$ time algorithm for testing if a given vertex set is a PMC~\cite{DBLP:journals/siamcomp/BouchitteT01}.

\begin{myproposition}[\cite{DBLP:journals/siamcomp/BouchitteT01}]
\label{pro:pmc_blocks}
If $\pcl$ is a PMC of a graph $G$ then all components $C \in \CC(G \setminus \pcl)$ are blocks of $G$.
\end{myproposition}

We call the components $C \in \CC(G \setminus \pcl)$ the blocks of $\pcl$.
Note that $N(C) \subsetneq \pcl$, i.e., the minimal separators of the blocks $C \in \CC(G \setminus \pcl)$ are strict subsets of the PMC.
We also need the following proposition connecting PMCs and blocks.

\begin{myproposition}[\cite{DBLP:journals/siamcomp/BouchitteT01}]
\label{pro:pmc_block2}
If $\pcl$ is a PMC of a graph $G$ and $C$ is a block of $\pcl$, then there is a full component $C'$ of $N(C)$ such that $\pcl \subseteq N[C']$.
\end{myproposition}

\begin{figure}[!t]
\centering
\begin{tikzpicture}[scale=1.5, every node/.style={draw, circle, minimum size=15pt, text centered,inner sep=0pt}]
\node (a) at (0,0.5) {$a$};
\node (b) at (-0.4,0) {$b$};
\node (c) at (0.4,0) {$c$};
\node (d) at (-0.8,-0.5) {$d$};
\node (e) at (0,-0.5) {$e$};
\node (f) at (0.8,-0.5) {$f$};

\path (a) edge (b);
\path (a) edge (c);
\path (b) edge (c);
\path (b) edge (d);
\path (c) edge (f);
\path (d) edge (e);
\path (e) edge (f);
\end{tikzpicture}
\caption{An example graph $G$ with vertex set $V(G) = \{a,b,c,d,e,f\}$. It holds that $\pcl = \{b,c,e\}$ is a PMC of $G$, and therefore by Proposition~\ref{pro:pmc_blocks}, $\{a\}$, $\{d\}$, and $\{f\}$ are blocks of $\pcl$.
As an example of Proposition~\ref{pro:pmc_block2}, consider $\pcl$ and the block $C = \{a\}$. Now $N(C) = \{b,c\}$ is a minimal separator contained in $\pcl$, and $C' = \{d,e,f\}$ is a full component of $\{b,c\}$.
It holds that $\pcl \subseteq N[C'] = \{b,c,d,e,f\}$.}
\label{fig:pmc_ex}
\end{figure}

\subsection{Formal Definitions of Problems\label{subsec:defs}}
Let $G$ be a graph and $\Tr(G)$ the set of triangulations of $G$.
The treewidth of $G$ is $\min_{H \in \Tr(G)} \max_{\pcl \in \MC(H)} |\pcl| -1$.
Given a weight function $w : V(G)^2 \rightarrow \mathbb{R}_{\ge 0}$, the weighted minimum fill-in of $G$ with respect to $w$ is $\min_{H \in \Tr(G)} \sum_{e \in (E(H) \setminus E(G))} w(e)$.
In unweighted minimum fill-in $w(e) = 1$.
When a set $F \subseteq V(G)^2 \setminus E(G)$ of admissible edges is given, the chordal sandwich problem is to determine if there is a triangulation $H \in \Tr(G)$ with $E(H) \subseteq E(G) \cup F$.
Note that chordal sandwich can be reduced to weighted minimum fill-in on the same graph $G$ by using a weight function $w$ with $w(e) = 0$ if $e \in F$ and $w(e) = 1$ otherwise.

A hypergraph $\mathcal{G}$ has a set of vertices $V(\mathcal{G})$ and a set of hyperedges $E(\mathcal{G})$ that are arbitrary non-empty subsets of $V(\mathcal{G})$, i.e., for $e \in E(\mathcal{G})$ it holds that $e \subseteq V(\mathcal{G})$.
The primal graph $P(\mathcal{G})$ of $\mathcal{G}$ is a graph with vertices $V(P(\mathcal{G})) = V(\mathcal{G})$ and edges $E(P(\mathcal{G})) = \bigcup_{e \in E(\mathcal{G})} e^2$.
An edge cover of a vertex set $X \subseteq V(\mathcal{G})$ is an assignment $c : E(\mathcal{G}) \rightarrow \{0, 1\}$ so that for each $v \in X$ it holds that $\sum_{v \in e \in E(\mathcal{G})} c(e) \ge 1$.
The size of an edge cover $c$ is $\sum_{e \in E(\mathcal{G})} c(e)$.
Fractional edge cover is defined analogously to edge cover, but the function $c$ is allowed to take any non-negative real values instead of only integers.
The minimum size of an edge cover of a set $X \subseteq V(\mathcal{G})$ is denoted by COV$(X)$ and fractional edge cover by FCOV$(X)$.
The generalized hypertreewidth of a hypergraph $\mathcal{G}$ is $\min_{H \in \Tr(P(\mathcal{G}))} \max_{\pcl \in \MC(H)} \text{COV}(\pcl)$ and the fractional hypertreewidth is $\min_{H \in \Tr(P(\mathcal{G}))} \max_{\pcl \in \MC(H)} \text{FCOV}(\pcl)$~\cite{DBLP:journals/jacm/GottlobMS09,DBLP:journals/talg/GroheM14,DBLP:journals/ipl/MollTT12}.
Note that FCOV$(X)$ can be computed in polynomial time by linear programming, but computing COV$(X)$ corresponds to the NP-complete set cover problem~\cite{DBLP:journals/talg/GroheM14}.

For all of the aforementioned problems there is an optimal solution corresponding to a minimal triangulation.
Furthermore, all of the problems except generalized hypertreewidth can be solved in $O^*(|\PC(G)|)$ time when $\PC(G)$ is given as an input~\cite{DBLP:journals/siamcomp/FominKTV08,DBLP:journals/tcs/FuruseY14,DBLP:conf/lata/Gysel14,DBLP:journals/dam/Lokshtanov10,DBLP:journals/ipl/MollTT12}.
Generalized hypertreewidth can be solved in $O^*(|\PC(P(\mathcal{G}))|)$ time if COV$(\pcl)$ for each $\pcl \in \PC(P(\mathcal{G}))$ is also given as input~\cite{DBLP:journals/ipl/MollTT12}.
For reductions from perfect phylogeny to chordal sandwich and from maximum compatibility of binary phylogenetic characters to weighted minimum fill-in see~\cite{DBLP:journals/dm/BordewichHS05,DBLP:conf/lata/Gysel14}.

\subsection{Notation on Edge Clique Cover}
We introduce notation for manipulating objects in a graph based on edge clique cover, with examples of the notation presented in Figure~\ref{fig:ecc_example}.
An edge clique cover $\w$ of a graph $G$ is a collection of subsets of $V(G)$ so that $\bigcup_{W \in \w} W^2 = E(G)$.
We often manipulate vertex sets based on an edge clique cover $\w$.
For a vertex $v \in V(G)$, we denote by $\w[v] = \{W \in \w \mid v \in W\}$ the set of cliques in $\w$ that contain $v$.
Similarly, for a vertex set $X$ we denote by $\w[X] = \bigcup_{v \in X} \w[v]$ the set of cliques in $\w$ that intersect $X$.

\begin{figure}
\centering
\begin{tikzpicture}[scale=1.5, every node/.style={draw, circle, minimum size=15pt, text centered,inner sep=0pt}]
\node (a) at (0,1) {xy};
\node (b) at (0.866,0.5) {y};
\node (c) at (0.866,-0.5) {yz};
\node (d) at (0,-1) {z};
\node (e) at (-0.866,-0.5) {xz};
\node (f) at (-0.866,0.5) {x};
\node (g) at (0,0) {xyz};

\path (a) edge (b);
\path (a) edge (g);
\path (b) edge (g);
\path (b) edge (c);
\path (a) edge (c);
\path (g) edge (c);

\path (a) edge (f);
\path (a) edge (e);
\path (f) edge (g);
\path (e) edge (f);
\path (e) edge (g);

\path (e) edge (c);
\path (g) edge (d);
\path (c) edge (d);
\path (e) edge (d);
\end{tikzpicture}
\caption{An example graph $G$ with vertex set $V(G) = \{x,y,z,xy,xz,yz,xyz\}$ and edge clique cover $\w = \{X,Y,Z\}$, where $X = \{x,xy,xz,xyz\}$, $Y = \{y,xy,yz,xyz\}$, and $Z = \{z,xz,yz,xyz\}$. Here, for example $\w[xy] = \{X,Y\}$, $\w[\{xy,xz\}] = \{X,Y,Z\}$, $V(G,\{X,Y\}) = \{x,y,xy\}$, $V(G, \{X,Y\}, \{Z\}) = \{x,y,xy,z\}$. Also, $\CC(\{X,Y\}) = \{\{x,y,xy\}\}$, and because $\{x,y,xy\}$ is a block, the part $\{X,Y\}$ is good. The parts $\{X\}$ and $\{Y\}$ are not compatible because $V(G,\{X\},\{Y\}) = \{x,y\}$ is not equal to $V(G, \{X,Y\}) = \{x,y,xy\}$.\label{fig:ecc_example}}
\end{figure}

A non-empty subset $\w' \subseteq \w$ of an edge clique cover $\w$ is called a \emph{part} of the edge clique cover.
The vertices in $V(G, \w') = \{v \in V(G) \mid \w[v] \subseteq \w'\}$ are called the vertices of the part $\w'$.
We denote the union of vertices of multiple parts with a shorthand $V(G, \w_1, \ldots, \w_p) = V(G, \w_1) \cup \ldots \cup V(G, \w_p)$.
The components of a part are $\CC(\w') = \CC(G[V(G, \w')])$.
A part is called \emph{good} if all of its components are blocks, in which case the components of the part may be called the blocks of the part.
Note that a part $\w'$ with $V(G, \w') = \emptyset$ has $\CC(\w') = \emptyset$ and thus is good.
Two disjoint parts $\w_1, \w_2$ are called \emph{compatible} if $V(G, \w_1, \w_2) = V(G, \w_1 \cup \w_2)$.

\section{Characterization of the Central Combinatorial Objects}
\label{sec:ccco}
In this section we show that if a graph has an edge clique cover of size $\ecc$, then the number of blocks of the graph is at most $S(\ecc, 2) \cdot 2$, the number of minimal separators is at most $S(\ecc, 2)$, and the number of potential maximal cliques is at most $S(\ecc, 3) + \ecc$.
The characterizations of these objects will be later used in the design of the algorithms.

We start by showing that blocks correspond to parts of an edge clique cover.

\begin{mylemma}
\label{lem:block_char}
Let $G$ be a graph and $\w$ an edge clique cover of $G$.
If $C$ is a block of $G$ then $V(G, \w[C]) = C$.
\end{mylemma}
\begin{proof}
Clearly $C \subseteq V(G, \w[C])$.
Note that $V(G, \w[C]) \subseteq N[C]$ because any vertex $v$ intersecting a common clique with a vertex $u \in C$ must be in $N[u]$.
Suppose there is a vertex $v \in (V(G, \w[C]) \cap N(C))$.
Let $C'$ be a full component of $N(C)$ distinct from $C$, implying that $v \in N(C')$.
All the cliques that $v$ intersects also intersect with $C$, and therefore there must be a vertex in $C'$ that is in a clique intersecting with $C$ which is a contradiction to the fact that $N(C)$ separates $C$ and $C'$.
\end{proof}

By Lemma~\ref{lem:block_char}, any block $C$ of $G$ can be uniquely identified with a set $\w[C] \subseteq \w$.
Moreover, there is no block with $\w[C] = \emptyset$ or $\w[C] = \w$, so the number of blocks is at most $2^\ecc -2$.
Any minimal separator is identified as $N(C)$ of at least two blocks $C$, so the number of minimal separators is at most $\frac{2^\ecc -2}{2} = 2^{\ecc-1} - 1 = S(\ecc, 2)$, proving the bound on minimal separators in Theorem~\ref{the:numbers}.

Next we show that each PMC $\pcl$ is either a clique in the edge clique cover $\w$ or can be represented by a tripartition of edge clique cover.
The high-level idea of the proof is that we can associate for each block $C \in \CC(G \setminus \pcl)$ a part $\w[C]$, and then use the properties of PMCs to show that the set of these parts can be coarsed into the desired tripartition, except in the special case when $\pcl \in \w$. 
The lemma could be a bit simpler if we wanted to only prove a bound of $3^\ecc$, but now it yields a tight bound.

\begin{mylemma}
\label{lem:pmc_char}
Let $G$ be a graph and $\w$ an edge clique cover of $G$.
If $\pcl$ is a PMC of $G$, then either (1) $\pcl \in \w$ or (2) $\CC(G \setminus \pcl) = \CC(\w_1) \cup \CC(\w_2) \cup \CC(\w_3)$, where $\{\w_1, \w_2, \w_3\}$ is a partition of $\w$ into good parts.
\end{mylemma}
\begin{proof}
First, we show that if there is a vertex $v \in \pcl$ with $|\w[v]| = 1$, then $\pcl = N[v]$, i.e., the sole clique containing $v$.
Consider a vertex $u \in N(v)$ and suppose that $u \notin \pcl$ for the sake of contradiction.
Note that $\w[v] \subseteq \w[u]$ and thus $N[v] \subseteq N[u]$.
Now, as $\pcl$ satisfies the cliquish condition there is a path from $v$ without intermediate vertices in $\pcl$ to all vertices in $\pcl$.
We can substitute $u$ for $v$ in any such path, and therefore $\pcl$ would violate the no full component condition.
Therefore we have $N[v] \subseteq \pcl$, which implies $N[v] = \pcl$ because now $v$ cannot satisfy the cliquish condition with any vertex not in $N[v]$.

We showed that case~1 applies to PMCs containing vertices $v$ with $|\w[v]| = 1$. 
Next, we suppose that we have PMC $\pcl$ for which case~1 does not apply, and show that case~2 applies.
By Proposition~\ref{pro:pmc_blocks} the components $C \in \CC(G \setminus \pcl)$ are blocks and by Lemma~\ref{lem:block_char} they define a collection $P = \{\w[C] \mid C \in \CC(G \setminus \pcl)\}$ of disjoint good parts of $\w$.
We make this collection a partition of $\w$ by adding the missing elements $\w \setminus (\bigcup_{\w_i \in P} \w_i)$ as singletons.
If for any such singleton $\{W\}$ there is a vertex $v$ with $\w[v] \subseteq \{W\}$, then case~1 applies.
Therefore we have that $V(G, \{W\}) = \emptyset$ for such singletons, and thus they are good parts.
Now we have a partition $P$ of $\w$ into good parts with $\bigcup_{\w_i \in P} \CC(\w_i) = \CC(G \setminus \pcl)$.

If the partition $P$ would consist of only a single part then $\pcl$ would be empty.
Suppose the number of parts is two, i.e., $P = \{\w_1, \w_2\}$.
Now, if either $V(G, \w_1)$ or $V(G, \w_2)$ is nonempty, then the no full component condition would be violated by it.
If both $V(G, \w_1)$ and $V(G, \w_2)$ are empty, then $\pcl = V(G)$, which implies that $\w_1$ and $\w_2$ are singletons and $|\w|=2$, which implies that $V(G) = \w_1 = \w_2 = \pcl$, i.e., that case~1 would apply.

If the number of parts is at least three, merge arbitrary compatible pairs of parts until the number of parts is three or no pairs of parts can be merged anymore.
If we end up with three parts, we are done.
If we end up with more than three parts, i.e. $P = \{\w_1, \w_2, \w_3, \w_4, \ldots\}$, then take a vertex $u \in (V(G, \w_1 \cup \w_2) \setminus V(G, \w_1, \w_2))$ and a vertex $v \in (V(G, \w_3 \cup \w_4) \setminus V(G, \w_3, \w_4))$.
Because of our assumption that we cannot continue the merging process anymore both of these vertices exist and are in $\pcl$.
However, there is no edge between $u$ and $v$ and there is no common component in whose neighborhood $u$ and $v$ are, which is a contradiction to the cliquish condition.
\end{proof}

We call the PMCs corresponding to case~1 of Lemma~\ref{lem:pmc_char} type~1 PMCs and the PMCs corresponding to case~2 type~2 PMCs.
The number of type~1 PMCs is at most $\ecc$.
Type~2 PMCs correspond to tripartitions of $\w$, so their number is at most $S(\ecc, 3)$.
Therefore, the total number of PMCs is at most $S(\ecc, 3) + \ecc$, completing the proof of Theorem~\ref{the:numbers}.

\section{Enumeration Algorithms}
\label{sec:enum_alg}
We modify the Bouchitt{\'{e}}--Todinca algorithm~\cite{DBLP:journals/tcs/BouchitteT02} for enumerating PMCs to give an $O^*(3^\ecc)$ time PMC enumeration algorithm and a polynomial space $O^*(9^\ecc)$ time PMC enumeration algorithm.
Recall that Theorem~\ref{the:numbers} with Takata's algorithm~\cite{DBLP:journals/dam/Takata10} implies a polynomial space $O^*(2^\ecc)$ time minimal separator enumeration algorithm.

The Bouchitt{\'{e}}--Todinca algorithm characterizes PMCs based on minimal separators.
We summarize this characterization in the following proposition.
\begin{myproposition}[\cite{DBLP:journals/tcs/BouchitteT02}]
\label{pro:bt_pmc_enum}
Let $G$ be a connected graph with $|V(G)| > 1$ and $v$ any vertex of $G$.
If $\pcl$ is a PMC of $G$, one of the following holds.
\begin{enumerate}
\item $\pcl \setminus \{v\} \in \PC(G \setminus \{v\})$.
\item $\pcl \setminus \{v\} \in \MS(G)$.
\item $\pcl = S \cup T$, where $S \in \MS(G)$ and $T \in \MS(G[C \cup \{x, y\}])$, where $C$ is a full component of $S$ and $x$ and $y$ are non-adjacent vertices in $S$.
\end{enumerate}
\end{myproposition}

The algorithm uses case~1 to generate $n$ induced subgraphs of the input graph, from which PMCs are generated by cases~2 and~3.
Note that the size of a minimum edge clique cover is monotone with respect to induced subgraphs.
We need the following lemma, which follows from Proposition~\ref{pro:pmc_basic}, to ensure that each PMC of each induced subgraph corresponds to at most one PMC of the original graph.

\begin{mylemma}[\cite{DBLP:journals/tcs/BouchitteT02}]
\label{lem:pmc_ind}
Let $G$ be a graph, $\pcl \in \PC(G)$, and $v \in \pcl$.
The set $\pcl \setminus \{v\}$ is not a PMC of $G$.
\end{mylemma}
\begin{proof}
The PMC $\pcl$ satisfies the cliquish condition, so there is a path in $G$ from $v$ to each vertex in $\pcl \setminus \{v\}$ without intermediate vertices in $\pcl$.
Therefore $\pcl \setminus \{v\}$ violates the no full component condition.
\end{proof}

Now, we can just enumerate PMCs from cases~2 and~3 in each of the $n$ induced subgraphs, and each time a PMC is found we use case~1 with Lemma~\ref{lem:pmc_ind} to generate at most one PMC of the original graph in polynomial time.

Next we complete the description of the algorithm by showing that PMCs from cases~2 and~3 can be enumerated in $O^*(3^\ecc)$ time.
The proof is based on Lemma~\ref{lem:block_char} on the structure and the number of blocks.

\begin{mylemma}
\label{lem:pmc_enum}
There is an algorithm that given a graph $G$ whose minimum edge clique cover has size $\ecc$ enumerates the PMCs of $G$, possibly with duplicates, in polynomial space and $O^*(3^\ecc)$ time.
\end{mylemma}
\begin{proof}
As discussed above, it is sufficient to enumerate PMCs from cases~2 and~3 of Proposition~\ref{pro:bt_pmc_enum}.
Case~2 simply corresponds to minimal separator enumeration, so we use the $O^*(2^\ecc)$ time polynomial space minimal separator enumeration algorithm.
For case~3, we do the polynomial space minimal separator enumeration in the graph $G$, and every time we output a minimal separator $S$, we do polynomial space enumeration of minimal separators in the graph $G[C \cup \{x, y\}]$ for all full components $C$ of $S$ and all non-adjacent pairs $x,y \in S$.

We do the inner iteration $O(n^2)$ times for each block $C$ of $G$.
The complexity of the inner iteration depends on the size of a minimum edge clique cover of the graph $G[C \cup \{x, y\}]$.
Let $\w$ be a minimum edge clique cover of $G$.
By Lemma~\ref{lem:block_char}, the block $C$ corresponds to a unique subset $\w[C]$ of $\w$.
The subset $\w[C]$ is an edge clique cover of $G[C \cup \{x, y\}]$ because all edges in it are adjacent to $C$ because $x$ and $y$ are non-adjacent.
Therefore, the time complexity of the inner iteration is $O^*(2^{|\w[C]|})$ and the total time complexity of the algorithm is $\sum_{\w' \subseteq \w} O^*(2^{|\w'|}) = O^*(3^\ecc)$.
\end{proof}

Using for example sorting we can deduplicate the output of the algorithm of Lemma~\ref{lem:pmc_enum} and an $O^*(3^\ecc)$ time and space algorithm for enumerating PMCs without duplicates, i.e., Theorem~\ref{the:pmc_enum}, follows.
For deduplication in polynomial space, we use a simple trick that is efficient enough for our purposes.

\begin{mylemma}
\label{lem:pmc_enum_polyspace}
There is an algorithm that given a graph $G$ whose minimum edge clique cover has size $\ecc$ enumerates the PMCs of $G$ in polynomial space and $O^*(9^\ecc)$ time.
\end{mylemma}
\begin{proof}
Run the algorithm of Lemma~\ref{lem:pmc_enum} multiple times in succession, each time outputting the lexicographically smallest PMC that is lexicographically larger than the previous PMC outputted, until no such PMC is found.
Now, using the algorithm at most $3^\ecc$ times we have outputted the PMCs of $G$ in lexicographically strictly increasing order.
\end{proof}

\section{Generalized Hypertreewidth}
\label{sec:ghtw}
For generalized hypertreewidth we need to compute minimum edge covers of all PMCs, so a naive application of Theorem~\ref{the:pmc_enum} with an $O^*(2^m)$ time set cover algorithm results in an $O^*(6^m)$ time algorithm, where $m$ is the number of hyperedges.
In this section we give an $O^*(4^m)$ time algorithm for enumerating PMCs with their minimum edge covers, implying an $O^*(4^m)$ time algorithm for generalized hypertreewidth.
We note that the naive approach with Lemma~\ref{lem:pmc_enum_polyspace} is sufficient to give an $O^*(9^m)$ time polynomial space algorithm for enumerating PMCs with minimum edge covers.

In this section we let $\w$ be the edge clique cover of size $m$ formed by the hyperedges of the input hypergraph.
We use the characterization of PMCs given in Lemma~\ref{lem:pmc_char}.
PMCs of type~1 have edge covers of size 1, so we focus on PMCs of type~2, which correspond to tripartitions of hyperedges.
Let $\{\w_1, \w_2, \w_3\}$ be a tripartition corresponding to a type~2 PMC $\pcl = V(G) \setminus V(G, \w_1, \w_2, \w_3)$.
We use the following notation to cover vertices of the PMC with hyperedges.

\begin{mydefinition}
\label{def:gam}
Let $G$ be a graph, $\w$ an edge clique cover of $G$, and $\{D, \w_1, \w_2, \w_3\}$ a labeled four-partition of $\w$ with empty parts allowed.
The partial PMC induced by $\{D, \w_1, \w_2, \w_3\}$ is the vertex set $\{v \in V(G) \mid D \cap \w[v] = \emptyset\} \setminus V(G, \w_1, \w_2, \w_3)$, which we denote by $\Gamma(G, D, \w_1, \w_2, \w_3)$.
\end{mydefinition}

Now we can observe that if a tripartition $\{\w_1, \w_2, \w_3\}$ corresponds to a PMC $\pcl$, then $\Gamma(G, \emptyset, \w_1, \w_2, \w_3) = \pcl$.
Furthermore, $\Gamma(G, D, \w_1 \setminus D, \w_2 \setminus D, \w_3 \setminus D)$ is an empty set if and only if $D$ is an edge cover of $\pcl$.
Therefore the task of finding a minimum edge cover of a PMC $\pcl = V(G) \setminus V(G, \w_1, \w_2, \w_3)$ reduces to finding a minimum size set $D \subseteq \w$ that makes $\Gamma(G, D, \w_1 \setminus D, \w_2 \setminus D, \w_3 \setminus D)$ empty.
We solve this with dynamic programming in $O^*(4^m)$ total time.

\begin{mylemma}
Let $\mathcal{G}$ be a hypergraph.
The PMCs of the primal graph $P(\mathcal{G})$ with minimum edge covers COV$(\pcl)$ for each PMC $\pcl$ can be enumerated in $O^*(4^m)$ time, where $m$ is the number of hyperedges of $\mathcal{G}$.
\end{mylemma}
\begin{proof}
Let $G = P(\mathcal{G})$, $\w = E(\mathcal{G})$, and $\{D, \w_1, \w_2, \w_3\}$ be as in Definition~\ref{def:gam}.
We denote by COV$(D, \w_1, \w_2, \w_3)$ the minimum size of $D' \subseteq \w$ such that the set $\Gamma(G, D \cup D', \w_1 \setminus D', \w_2 \setminus D', \w_3 \setminus D')$ is empty.
We compute COV for all such partitions by the recursion $\text{COV}(D, \w_1, \w_2, \w_3) =$
$$
\left\{
	\begin{array}{l}
		0 \mbox{  if } \Gamma(G, D, \w_1, \w_2, \w_3) = \emptyset \mbox{ and } \\
		1 + \text{min}_{W \in \w} \text{COV}(D \cup \{W\}, \w_1 \setminus \{W\}, \w_2 \setminus \{W\}, \w_3 \setminus \{W\}) \mbox{  otherwise.} \\
	\end{array}
\right.$$
This recursion can be evaluated and stored in $O^*(4^m)$ time.
Now, we iterate over all tripartitions $\{\w_1, \w_2, \w_3\}$ of $\w$ and if $V(G) \setminus V(G, \w_1, \w_2, \w_3)$ is a PMC we output it and COV$(\emptyset, \w_1, \w_2, \w_3)$ as its minimum edge cover.
\end{proof}

Theorem~\ref{the:ghtw} follows.

\section{Faster Algorithms When Edge Clique Cover is Given}
\label{sec:fast}
We use fast subset convolution~\cite{DBLP:conf/stoc/BjorklundHKK07} to design $O^*(2^{\ecc'})$ time algorithms for treewidth, minimum fill-in, and chordal sandwich, where $\ecc'$ is the size of an edge clique cover given as an input.
In particular, we make use of the following result.

\begin{myproposition}[\cite{DBLP:conf/stoc/BjorklundHKK07}]
Let $X$ be a set, $f : 2^X \to [M]$ and $g : 2^X \to [M]$ functions from the set $2^X$ of all subsets of $X$ to the set of integers up to $M$.
The function $(f * g)$ defined as $(f * g)(Y) = \min_{Y' \subseteq Y} f(Y') + g(Y \setminus Y')$ can be computed for all subsets $Y \subseteq X$  in $O^*(2^{|X|} M)$ time.
\end{myproposition}

The algorithms we introduce are modifications of the dynamic programming phase of the PMC framework.
In most of this section our presentation is general in the sense that it applies to each of the three problems.
We use the term ``optimal triangulation'' to refer to a triangulation with the minimum size of a maximum clique in the context of treewidth, a triangulation with the least number of edges in the context of minimum fill-in, and to a triangulation that has no non-admissible fill-edges in the context of chordal sandwich (or to information that no such triangulation exists).

We start by recalling the dynamic programming phase of the PMC framework.
The states of the dynamic programming correspond to \emph{realizations} of blocks.
\begin{mydefinition}[\cite{DBLP:journals/siamcomp/BouchitteT01}]
Let $G$ be a graph and $C$ a block of $G$.
A realization $R(C)$ of $C$ is a graph with vertex set $V(R(C)) = N[C]$ and edge set $E(R(C)) = E(G[N[C]]) \cup N(C)^2$.
\end{mydefinition}

The following proposition characterizes minimal triangulations of a realization of a block.

\begin{myproposition}[\cite{DBLP:journals/siamcomp/BouchitteT01}]
\label{pro:bt_rec}
Let $G$ be a graph and $C$ a block of $G$.
The graph $H$ is a minimal triangulation of $R(C)$ if and only if (i) $V(H) = N[C]$ and (ii) there is a PMC $\pcl \in \PC(G)$ with $N(C) \subseteq \pcl \subseteq N[C]$ and
$$E(H) = \pcl^2 \cup \bigcup_{C_i \in \CC(R(C) \setminus \pcl)} E(H_i),$$
where $H_i$ is any minimal triangulation of $R(C_i)$.
Moreover, each $C_i$ is a block of $G$.
\end{myproposition}

Proposition~\ref{pro:bt_rec} implies dynamic programming algorithms for computing optimal triangulations of realizations of all blocks~\cite{DBLP:journals/siamcomp/BouchitteT01,DBLP:journals/siamcomp/FominKTV08,DBLP:journals/dam/Lokshtanov10}.

We use the following proposition for a base case.

\begin{myproposition}[\cite{DBLP:journals/siamcomp/BouchitteT01}]
\label{pro:bt_base}
Let $G$ be a graph that is not complete.
The graph $H$ is a minimal triangulation of $G$ if and only if (i) $V(H) = V(G)$ and (ii) there is a minimal separator $S \in \MS(G)$ with
$$E(H) = \bigcup_{C_i \in \CC(G \setminus S)} E(H_i),$$
where $H_i$ is any minimal triangulation of $R(C_i)$.
\end{myproposition}

Once we have computed optimal triangulations of realizations of all blocks, we can compute an optimal triangulation of the graph via Proposition~\ref{pro:bt_base} in time $O^*(2^\ecc)$.
For computing optimal triangulations of realizations, the bottleneck in implementing the recursion of Proposition~\ref{pro:bt_rec} is in iterating over the PMCs.

The high-level idea of our algorithm is that we use Proposition~\ref{pro:bt_rec} directly only with type~1 PMCs, i.e., PMCs $\pcl \in \w \cap \PC(G)$.
For type~2 PMCs, we simulate the iteration over PMCs with fast subset convolution.
In particular, we show that each PMC $\pcl$ of type~2 with $N(C) \subseteq \pcl \subseteq N[C]$ can be expressed in terms of two disjoint good parts $\w_1$ and $\w_2$ of $\w[C]$, where $\w$ is an edge clique cover of $G$.
In the case of treewidth, minimum fill-in, and chordal sandwich, an optimal partition of every subset $\w' \subseteq \w$ into two good parts $\w_1$ and $\w_2$ can be computed with fast subset convolution, provided that we have first computed optimal triangulations of realizations of all blocks in $\CC(\w_1)$ and $\CC(\w_2)$.

We first show that each type~2 PMC can be expressed in terms of $\w_1$ and $\w_2$.

\begin{mylemma}
\label{lem:pmc_fast1}
Let $G$ be a graph, $\w$ an edge clique cover of $G$, and $C$ a block of $G$.
If a graph $H$ is a minimal triangulation of $R(C)$, then either (1) there is a clique $\pcl \in \w \cap \PC(G)$ and
$$E(H) = \pcl^2 \cup \bigcup_{C_i \in \CC(R(C) \setminus \pcl)} E(H_i),$$
where $N(C) \subseteq \pcl \subseteq N[C]$, and $H_i$ is a minimal triangulation of $R(C_i)$, or (2) there is a partition $\{\w_1, \w_2, \w_o\}$ of $\w$ into good parts with $\w_1 \cup \w_2 \subseteq \w[C]$, a block $C' \in \CC(\w_o)$ with $N(C) = N(C')$, and
$$E(H) = \pcl^2 \cup \bigcup_{C_i \in \CC(\w_1) \cup \CC(\w_2) \cup (\CC(\w_o) \setminus \CC(G \setminus N(C)))} E(H_i),$$
where $\pcl = V(G) \setminus V(G, \w_1, \w_2, \w_o)$ and $H_i$ is a minimal triangulation of $R(C_i)$.
\end{mylemma}
\begin{proof}
Case~1 corresponds to Proposition~\ref{pro:bt_rec} with PMCs of type~1.
Next we prove that case~2 covers all PMCs of type~2.

Let $\pcl$ be a PMC of $G$ with $N(C) \subseteq \pcl \subseteq N[C]$ and $\pcl \notin \w$.
We consider the part $\w'_o = \w \setminus \w[C]$ and prove that $\w'_o$ is a good part and $\CC(\w'_o) = \CC(G \setminus N(C)) \setminus \{C\}$.
Observe that $\{C, N(C), V(G, \w'_o)\}$ is a partition of $V(G)$, and moreover there are no edges between $C$ and $V(G, \w'_o)$.
Now, for any component $C' \in \CC(\w'_o)$, it must hold that $N(C') \subseteq N(C)$, and therefore $N(C')$ is a minimal separator and therefore $\w'_o$ is a good part whose blocks are the components of $G \setminus N(C)$ except $C$.

Similarly as in the proof of Lemma~\ref{lem:pmc_char}, consider the collection of disjoint good parts $P = \{\w[C_i] \mid C_i \in \CC(G \setminus \pcl)\}$.
All of the parts that intersect $\w'_o$ are subsets of $\w'_o$ because they do not intersect $\w[C]$, and therefore we replace the parts that intersect $\w'_o$ by the part $\w'_o$.
Now by similar arguments as in Lemma~\ref{lem:pmc_char} we add additional parts to the collection to make it a full partition of $\w$.
Now we have a partition $P$ of $\w$ with $|P| \ge 2$ and $\w'_o \in P$.
Moreover, $\bigcup_{\w_i \in P} \CC(\w_i) = \CC(G \setminus \pcl)$.
If $|P| = 2$, then it would hold that $P = \{\w'_o, \w[C]\}$, in which case $\pcl = N(C)$ would hold, which is a contradiction.
If there are at least three parts, then merge compatible parts until we have three parts or cannot merge parts anymore.
By the proof of Lemma~\ref{lem:pmc_char}, we will end up with three parts.
Now let $\w_o$ be the part that contains $\w'_o$ and $\w_1$ and $\w_2$ the other two parts.
Because $\w'_o = \w \setminus \w[C]$, we have that $\w_1 \cup \w_2 \subseteq \w[C]$.
Moreover, because $N(C)$ has at least two full components, and all components of $N(C)$ except $C$ are components of $\w'_o$, we have that there is a block $C' \in \CC(\w_o)$ with $N(C') = N(C)$.

Finally, we show that $\CC(R(C) \setminus \pcl) = \CC(\w_1) \cup \CC(\w_2) \cup (\CC(\w_o) \setminus \CC(G \setminus N(C)))$.
By Proposition~\ref{pro:pmc_block2} we have that $\CC(R(C) \setminus \pcl) = \CC(G \setminus \pcl) \setminus \CC(G \setminus N(C))$ and therefore $\CC(R(C) \setminus \pcl) = \CC(\w_1) \cup \CC(\w_2) \cup \CC(\w_o) \setminus \CC(G \setminus N(C))$.
All blocks of $\w_1$ and $\w_2$ are subsets of $C$ because $\w_1 \cup \w_2 \subseteq \w[C]$.
\end{proof}

The following lemma guarantees that the characterization of PMCs of type~2 in Lemma~\ref{lem:pmc_fast1} is sound in the sense that all graphs $H$ that it defines are (not necessarily minimal) triangulations of $G$.

\begin{mylemma}
\label{lem:pmc_fast2}
Let $G$ be a graph, $\w$ an edge clique cover of $G$, and $C$ a block of $G$.
Also, let $\{\w_1, \w_2, \w_o\}$ be a partition of $\w$ into good parts with $\w_1 \cup \w_2 \subseteq \w[C]$ and $C'$ a block of $G$ with $C' \in \CC(\w_o)$ and $N(C) = N(C')$.
Let $H$ be any graph with (i) $V(H) = N[C]$ and (ii)
$$E(H) = \pcl^2 \cup \bigcup_{C_i \in \CC(\w_1) \cup \CC(\w_2) \cup (\CC(\w_o) \setminus \CC(G \setminus N(C)))} E(H_i),$$
where $\pcl = V(G) \setminus V(G, \w_1, \w_2, \w_o)$ and $H_i$ is any triangulation of $R(C_i)$.
The graph $H$ is a triangulation of $R(C)$.
\end{mylemma}
\begin{proof}
To show that $H$ is a triangulation of $R(C)$ we show that $N(C) \subseteq \pcl \subseteq N[C]$ and $\CC(R(C) \setminus \pcl) = \CC(\w_1) \cup \CC(\w_2) \cup (\CC(\w_o) \setminus \CC(G \setminus N(C)))$.

For $\pcl \subseteq N[C]$, note that no vertex of $\pcl$ is in $V(G, \w_o)$, and therefore all vertices of $\pcl$ intersect a clique in $\w[C]$, and therefore are in $N[C]$.
Furthermore, $N(C) \subseteq \pcl$ holds because there is a block $C' \in \CC(\w_o)$ with $N(C) = N(C')$.

The components of $\pcl$ in $G$ are $\CC(\w_1) \cup \CC(\w_2) \cup \CC(\w_o)$ by definition.
Because $N(C) \subseteq \pcl \subseteq N[C]$, all components of $\pcl$ in $G$ are either in $C$, and thus components of $\pcl$ in $R(C)$ or are components of $N(C)$.
All components in $\CC(\w_1) \cup \CC(\w_2)$ are subsets of $C$ because $\w_1 \cup \w_2 \subseteq \w[C]$, and therefore are components of $\pcl$ in $R(C)$.
All components of $N(C)$ are outside of $C$ except $C$ itself.
It remains to show that $C$ is not a component of $\pcl$.
Any set $\w'$ with $C \subseteq V(G, \w')$ must be a superset of $\w[C]$.
However, because $\w_1$ and $\w_2$ are proper subsets of $\w[C]$, no set in $\{\w_1, \w_2, \w_o\}$ can be a superset of $\w[C]$.
\end{proof}

In Lemmas~\ref{lem:pmc_fast1} and~\ref{lem:pmc_fast2} we formulated a recursion that characterizes all minimal triangulations of a realization $R(C)$ of a block $C$ in terms of minimal triangulations of realizations $R(C')$ of blocks $C' \subsetneq C$.
What remains is to integrate the computation of the optimal cost of the triangulation into this characterization.
The following lemma is used for treewidth and minimum fill-in.
It is simple, but we state it as a warmup for what follows.

\begin{mylemma}
\label{lem:cost_tw}
Let $G$ be a graph, $\w$ an edge clique cover of $G$, $\{\w_1, \w_2, \w_o\}$ a partition of $\w$, and $\pcl = V(G) \setminus V(G, \w_1, \w_2, \w_o)$ a vertex set defined by the partition.
It holds that $|\pcl| = |V(G)| - |V(G, \w_1)| - |V(G, \w_2)| - |V(G, \w_o)|$.
\end{mylemma}
\begin{proof}
The sets $V(G, \w_1)$, $V(G, \w_2)$, and $V(G, \w_o)$ are disjoint because $\w_1$, $\w_2$ and $\w_o$ are disjoint.
\end{proof}
Therefore, the size of $\pcl$ can be computed as a sum that considers the parts $\w_1$, $\w_2$, and $\w_o$ independently, and therefore we can integrate the computation of it into fast subset convolution.
For treewidth, we only have to make sure that $|\pcl| \le k+1$, where $k$ is the upper bound for treewidth in the decision problem.
For minimum fill-in, we can compute the number of edges in the triangulation of $R(C)$ as $\binom{|\pcl|}{2} + \sum_{C_i \in \CC(R(C) \setminus \pcl)} (|E(H_i)| - \binom{|N(C_i)|}{2})$, where $H_i$ is an optimal triangulation of the realization $R(C_i)$.

A similar lemma is used for chordal sandwich.
\begin{mylemma}
\label{lem:cost_cs}
Let $G$ be a graph, $\w$ an edge clique cover of $G$, $\{\w_1, \w_2, \w_o\}$ a partition of $\w$ into good parts, and $\pcl = V(G) \setminus V(G, \w_1, \w_2, \w_o)$.
It holds that $\pcl^2 = \bigcup_{\w_i \in \{\w_1, \w_2, \w_o\}} (V(G) \setminus V(G, \w_i, \w \setminus \w_i))^2$.
\end{mylemma}
\begin{proof}
We first show that for all pairs $u,v \in \pcl$ there is a part $\w_i \in \{\w_1, \w_2, \w_o\}$ such that $\{u, v\} \subseteq (V(G) \setminus V(G, \w_i, \w \setminus \w_i))$.
Neither $\w[u]$ or $\w[v]$ is a subset of any of $\w_1$, $\w_2$, or $\w_o$, so there is at least of part of the partition that they both intersect but are not subsets of and that is the desired part $\w_i$.

To see that $(V(G) \setminus V(G, \w_i, \w \setminus \w_i)) \subseteq \pcl$, take any vertex $v \in (V(G) \setminus \pcl)$ and consider the set $\w[v]$.
The set $\w[v]$ is a subset of one of $\w_1$, $\w_2$, or $\w_o$.
Now, each of $\{\w_1, \w_2, \w_o\}$ is either $\w_i$ or a subset of $\w \setminus \w_i$.
\end{proof}

Lemma~\ref{lem:cost_cs} guarantees that each fill-edge caused by the PMC $\pcl$ can be ``seen'' from at least one of the parts $\w_1$, $\w_2$, $\w_o$, implying that it is sufficient to check each part independently to guarantee that $\pcl$ does not add any forbidden fill-edges.
We remark that Lemma~\ref{lem:cost_cs} appears to be difficult to generalize to count the exact number of fill-edges, which is the barrier why we are not able to give an $O^*(2^{\ecc'})$ time algorithm for weighted minimum fill-in.

Algorithm~\ref{alg:twfast} presents the full $O^*(2^{\ecc'})$ time algorithm for treewidth.
The algorithms for minimum fill-in and chordal sandwich are similar.
The algorithm maintains a collection $\mathcal{B}$ of blocks $C$ for which it is known that the treewidth of $R(C)$ is at most $k$.
The invariant of the main loop of lines~\ref{alg1:line:main_1} to~\ref{alg1:line:main_2} is that after $i$th iteration, all blocks of size at most $i$ and treewidth at most $k$ have been added to $\mathcal{B}$.
In each iteration of the main loop, the algorithm iterates over all good parts $\w'$ on lines~\ref{alg1:line:f11} to~\ref{alg1:line:f12}, and if all realizations of blocks of the part have treewidth at most $k$ adds the part to a collection $F_{|V(G, \w')|}$.
These parts $\w'$ correspond to parts $\w_1$ and $\w_2$ of our lemmas.
Then, fast subset convolution is applied on line~\ref{alg:line:fsc} on the collections $F$ to find for all combinations $(\w_1 \cup \w_2)$ of disjoint good parts $\w_1$ and $\w_2$ the maximum number of vertices in $V(G, \w_1, \w_2)$.
Then on lines~\ref{alg:line:wo1} to~\ref{alg:line:wo2} the algorithm iterates through all good parts $\w_o$, thus determining $(\w_1 \cup \w_2)$ and all other variables that need to be taken into account.
In particular, note that each part $\w_o$ determines only polynomially many blocks $C$ such that there is $C' \in \CC(\w_o)$ with $N(C') = N(C)$.

\begin{algorithm}[!t]
\caption{Treewidth in $O^*(2^{\ecc'})$ time}
\DontPrintSemicolon
\label{alg:twfast}
\SetKwInOut{Input}{Input}\SetKwInOut{Output}{Output}
\Input{Connected graph $G$, an edge clique cover $\w$ of $G$, and an integer $k$}
\Output{Whether the treewidth of $G$ is at most $k$}
\lIf{$G$ is complete}{
  \Return $|V(G)| \le k+1$
}
Let $\mathcal{B} \gets \emptyset$ be a collection of blocks of $G$\;
\For{$i \gets 1$ \KwTo $n$~\label{alg1:line:main_1}}{
  For each $0 \le j \le n$ let $F_j \gets \emptyset$ be a collection of subsets of $\w$\;
  \ForEach{good part $\w' \subseteq \w$~\label{alg1:line:f11}}{
    \If{$\CC(\w') \subseteq \mathcal{B}$}{
      $F_{|V(G, \w')|} \gets F_{|V(G, \w')|} \cup \{\w'\}$~\label{alg1:line:f12}\;
    }
  }
  Use fast subset convolution to compute for each subset $\w' \subseteq \w$ the maximum value of $a + b$ such that there is $\w'' \subseteq \w'$ with $\w'' \in F_a$ and $(\w' \setminus \w'') \in F_b$~\label{alg:line:fsc}\;
  \ForEach {good part $\w_o \subseteq \w$~\label{alg:line:wo1}}{
    Let $t$ be the value on $\w \setminus \w_o$ computed in line~\ref{alg:line:fsc}\;
    \If{$t$ exists and $n - t - |V(G, \w_o)| \le k+1$}{
      \ForEach {minimal separator $N(C')$ with $C' \in \CC(\w_o)$}{
        \If{exists $C \in \CC(G \setminus N(C'))$ with $(\w \setminus \w_o) \subseteq \w[C]$}{
          \If{$(\CC(\w_o) \setminus \CC(N(C))) \subseteq \mathcal{B}$}{
            $\mathcal{B} \gets \mathcal{B} \cup \{C\}$~\label{alg:line:wo2}\;
          }
        }
      }
    }
  }
  \ForEach {block $C$ of $G$}{
    \ForEach {$\pcl \in \w \cap \PC(G)$ with $|\pcl| \le k+1$ and $N(C) \subseteq \pcl \subseteq N[C]$}{
      \If{$\CC(R(C) \setminus \pcl) \subseteq \mathcal{B}$}{
        $\mathcal{B} \gets \mathcal{B} \cup \{C\}$~\label{alg1:line:main_2}\;
      }
    }
  }
}
\ForEach{$S \in \MS(G)$}{
  \If{$\CC(G \setminus S) \subseteq \mathcal{B}$}{
    \Return True\;
  }
}
\Return False\;
\end{algorithm}

The analysis of the algorithm focuses on transitions via PMCs of type~2 and proceeds by induction on the main loop invariant.
The time complexity follows simply from fast subset convolution, the bound $2^\ecc$ on the number of blocks, and the fact that each iteration of the loop of the lines~\ref{alg:line:wo1} to~\ref{alg:line:wo2} takes polynomial time.
The correctness is shown by combining the lemmas introduced in this section.

\begin{mylemma}
There is an algorithm that given a graph $G$ with an edge clique cover $\w$ of size $\ecc'$ determines the treewidth and minimum fill-in of $G$ in time $O^*(2^{\ecc'})$.
Furthermore, if also a set $F \subseteq V(G)^2 \setminus E(G)$ is given the algorithm determines if there is a triangulation $H$ of $G$ with $E(H) \subseteq E(G) \cup F$.
\end{mylemma}
\begin{proof}
By Theorem~\ref{the:numbers} and Proposition~\ref{pro:bt_base} it suffices to solve each of the problems for all realizations of blocks of $G$.
We proceed via induction, i.e., assume that the problems have been solved for all realizations of blocks $C$ with $|C| < i$.
We show that in one step taking a total of $O^*(2^{\ecc'})$ time the problems can be solved for all realizations of blocks $C$ with $|C| = i$.

First we find optimal triangulations via Proposition~\ref{pro:bt_rec} using PMCs of type~1, i.e., PMCs $\pcl \in \w \cap \PC(G)$.
For deciding if treewidth is at most $k$ we have to assert that $|\pcl| \le k+1$ and that the resulting smaller blocks have treewidth of realizations at most $k$.
For minimum fill-in, the number of fill-edges can be computed as $|\pcl^2 \setminus E(R(C))|$ plus the numbers of fill-edges in optimal solutions of the smaller blocks.
For chordal sandwich we have to assert that $\pcl^2 \subseteq E(G) \cup F$ and that the resulting smaller blocks also have positive answers.

Now we can focus on type~2 PMCs, i.e., implement the characterization of Lemmas~\ref{lem:pmc_fast1} and~\ref{lem:pmc_fast2} while keeping track of the optimal answer.
For each of the problems we use fast subset convolution~\cite{DBLP:conf/stoc/BjorklundHKK07} to determine in $O^*(2^{\ecc'})$ time for each subset $\w' \subseteq \w$ an optimal partition of $\w'$ into two good parts $\w_1$ and $\w_2$.
For treewidth we determine the maximum value of $|V(G, \w_1, \w_2)|$ such that all realizations of blocks in $\CC(\w_1) \cup \CC(\w_2)$ have treewidth at most $k$.
For minimum fill-in, we determine for each value $0 \le p \le n$ the minimum value of sum of minimum fill-ins of realizations of blocks in $\CC(\w_1) \cup \CC(\w_2)$ such that $|V(G, \w_1, \w_2)| = p$.
For chordal sandwich, we decide if there is a partition such that the answer is positive for all realizations of blocks in $\CC(\w_1) \cup \CC(\w_2)$ and that $(V(G) \setminus V(G, \w_i, \w \setminus \w_i))^2 \subseteq E(G) \cup F$.
Now, once we fix the set $\w' = (\w_1 \cup \w_2)$, by Lemmas~\ref{lem:cost_tw} and~\ref{lem:cost_cs} we have computed all relevant information about $\w_1$ and $\w_2$ with the convolution, and can compute the rest of the information of the PMC by $\w_o = \w \setminus (\w_1 \cup \w_2)$.
Because $\w_o$ defines polynomially many blocks $C$ with $C' \in \CC(\w_o)$ and $N(C) = N(C')$ we can update all relevant blocks in polynomial time.
\end{proof}

Theorem~\ref{the:fast} follows.

\section{Polynomial Space Algorithms}
\label{sec:poly}
We give polynomial space algorithms for treewidth, weighted minimum fill-in, and generalized and fractional hypertreewidth.
The algorithms are based on the following characterization of minimal triangulations.
Note that the characterization uses realizations of blocks as defined in Section~\ref{sec:fast}.

\begin{myproposition}[\cite{DBLP:journals/siamcomp/BouchitteT01}]
\label{pro:bt_rec_pmc}
Let $G$ be a graph, $H$ a minimal triangulation of $G$, and $\pcl$ a maximal clique of $H$.
For each $C_i \in \CC(G \setminus \pcl)$ there exists a minimal triangulation $H_i$ of $R(C_i)$ such that  
 $$E(H) = \pcl^2 \cup \bigcup_{C_i \in \CC(G \setminus \pcl)} E(H_i).$$
\end{myproposition}

Note that by iterating over all $\pcl \in \PC(G)$ in Proposition~\ref{pro:bt_rec_pmc} we can indeed construct all minimal triangulations of $G$.
Furthermore, all graphs $H$ constructed in this manner are minimal triangulations~\cite{DBLP:journals/siamcomp/BouchitteT01}.

The idea of the algorithm is to use the recursion of Proposition~\ref{pro:bt_rec_pmc} directly, without dynamic programming.
The following ``balanced PMC'' lemma guarantees that for any resulting minimal triangulation, we can construct it in an order in which the size of an edge clique cover roughly halves in each level of the recursion.
We prove it by an edge direction argument in a tree corresponding to the minimal triangulation.

\begin{mylemma}
\label{lem:balance}
Let $G$ be a graph with an edge clique cover $\w$.
Any minimal triangulation $H$ of $G$ has a maximal clique $\pcl$ so that all blocks $C \in \CC(G \setminus \pcl)$ have $|\w[C]| \le |\w|/2$.
\end{mylemma}
\begin{proof}
For any PMC $\pcl$ there can be at most one component $C \in \CC(G \setminus \pcl)$ so that $|\w[C]| > |\w|/2$ because the sets $\w[C_i]$ over $C_i \in \CC(G \setminus \pcl)$ correspond to disjoint subsets of $\w$.
Let $H$ be any minimal triangulation of $G$ and pick arbitrary maximal clique $\pcl$ of $H$.
While there is a component $C \in \CC(G \setminus \pcl)$ such that $|\w[C]| > |\w|/2$, pick a maximal clique $\pcl$ of $H$ such that $N(C) \subseteq \pcl \subseteq N[C]$.
If this process stops, we have found the desired maximal clique $\pcl$.
Suppose the process does not stop.
It considers an infinite sequence of blocks $C_1, C_2, \ldots$ with an associated infinite sequence of PMCs $\pcl_1, \pcl_2, \ldots$ with $C_i \in \CC(G \setminus \pcl_i)$.
Consider two consecutive blocks $C_i$ and $C_{i+1}$ in this sequence such that $C_{i+1}$ is not a subset of $C_i$, which exist because $G$ is finite.
Recall that $N(C_i) \subseteq \pcl_{i+1} \subseteq N[C_i]$.
Because $C_{i+1}$ is not a subset of $C_i$, we have that $N(C_{i+1}) \subseteq N(C_i)$, implying that $C_i$ and $C_{i+1}$ are two distinct components of $G \setminus N(C_i)$.
Therefore the sets $\w[C_i]$ and $\w[C_{i+1}]$ are disjoint, implying that either of them has to be of size at most $|\w|/2$, which is a contradiction.
\end{proof}

We combine Proposition~\ref{pro:bt_rec_pmc} and Lemma~\ref{lem:balance} into the following lemma.

\begin{mylemma}
\label{lem:tri_balance}
Let $G$ be a graph and $\w$ an edge clique cover of $G$.
A graph $H$ is a minimal triangulation of $G$ if and only if (1) $V(H) = V(G)$ and (2) there is a PMC $\pcl \in \PC(G)$ with $|\w[C_i]| \le |\w|/2$ for all $C_i \in \CC(G \setminus \pcl)$ and
$$E(H) = \pcl^2 \cup \bigcup_{C_i \in \CC(G \setminus \pcl)} E(H_i),$$
where $H_i$ is a minimal triangulation of $R(C_i)$.
\end{mylemma}
\begin{proof}
Such a graph $H$ is a minimal triangulation of $G$ by standard results~\cite{DBLP:journals/siamcomp/BouchitteT01}.
Let $H$ be any minimal triangulation of $G$.
By Lemma~\ref{lem:balance} there is a maximal clique $\pcl$ of $H$ that satisfies $|\w[C_i]| \le |\w|/2$ for all $C_i \in \CC(G \setminus \pcl)$.
By Proposition~\ref{pro:bt_rec_pmc} the triangulation $H$ can be constructed by the recursion from $\pcl$.
\end{proof}

\begin{algorithm}[!b]
\caption{Treewidth in polynomial space and $O^*(9^{\ecc'})$ time}
\DontPrintSemicolon
\label{alg:polyspace1}
\SetKwInOut{Input}{Input}\SetKwInOut{Output}{Output}
\Input{Connected graph $G$, an edge clique cover $\w$ of $G$, and an integer $k$}
\Output{Whether the treewidth of $G$ is at most $k$}
\For{$\pcl \in \PC(G)$}{
  \If{$|\pcl| \le k+1$ and $|\w[C_i]| \le |\w|/2$ for all $C_i \in \CC(G \setminus \pcl)$}{
    ok $\gets$ True\;
    \For{$C \in \CC(G \setminus \pcl)$}{
      \If{Treewidth$(R(C), \w[C] \cup \{N(C)\}, k)$ = False}{
        ok $\gets$ False\;
      }
    }
    \lIf{ok}{
      \Return True
    }
  }
}
\Return False\;
\end{algorithm}

Algorithm~\ref{alg:polyspace1} presents a polynomial space $O^*(9^{\ecc'})$ time algorithm for treewidth.
The algorithms for other problems are similar.
The algorithm implements the characterization of Lemma~\ref{lem:tri_balance}, with the observation that $\w[C] \cup \{N(C)\}$ is an edge clique cover of $R(C)$.
The time complexity analysis of the algorithm reduces to a recursion equation resembling $t(\ecc') = 9^{\ecc'} + 3^{\ecc'} 2 t({\ecc'}/2)$.

\begin{mylemma}
\label{lem:poly_alg1}
There is an algorithm that given a graph $G$ with an edge clique cover of size $\ecc'$ determines the treewidth of $G$ in polynomial space and $O^*(9^{\ecc'})$ time.
If also a weight function $w : V(G)^2 \rightarrow \mathbb{R}_{\ge 0}$ is given, the algorithm determines the weighted minimum fill-in of $G$ with respect to $w$.
There is also algorithm that given a hypergraph $\mathcal{G}$ with $m$ hyperedges determines both its generalized and fractional hypertreewidth in polynomial space and $O^*(9^m)$ time.
\end{mylemma}
\begin{proof}
We use a recursive procedure that takes as an input a graph $G$ and an edge clique cover $\w$ of it and returns the cost of an optimal triangulation of the graph.
We use the characterization of minimal triangulations of Lemma~\ref{lem:tri_balance} with the $O^*(9^\ecc)$ time polynomial space PMC enumeration algorithm of Lemma~\ref{lem:pmc_enum_polyspace}.
Note that if $C$ is a block, then $\w[C] \cup \{N(C)\}$ is an edge clique cover of $R(C)$.
Therefore, when we recurse into a subproblem the value of $\ecc'$ changes to at most $\ecc'/2+1$.

We use $k = \ecc'$ for clarity.
We analyze the time complexity of the algorithm by induction on $k$.
Our assumption is that the time complexity is $p(n) 9^{k+3 \log_2 k} k$, where $p(n)$ is a polynomial sufficiently large to cover all polynomial-time subroutines and to make the assumption true for small values of $k$.
In particular we assume that the PMCs can be enumerated and their associated costs can be computed in $p(n) 9^k$ time.
We recurse from at most $3^k$ PMCs $\pcl$, from each into $|\CC(G \setminus \pcl)|$ subproblems with the new value of $k$ being at most $k/2+1$.
Because of the exponentiality of the algorithm the worst case is that we recurse into two subproblems from each PMC, each subproblem with the new value of $k$ equal to $k/2+1$.
Now, the time complexity is 
\begin{align*}
&p(n) 9^k + 3^k 2 p(n) 9^{k/2+1+3 \log_2 (k/2+1)} (k/2+1)\\
=& p(n) 9^k + p(n) 9^{k+1+3\log_2(k+2)-3} (k+2)\\
\le& p(n) 9^k + p(n) 9^{k+3\log_2 k} (k+2)/9 \le p(n) 9^{k+3 \log_2 k} k,
\end{align*}
which verifies the induction assumption.
\end{proof}
Theorem~\ref{the:poly_alg1} follows.

\begin{algorithm}[!b]
\caption{Treewidth in polynomial space and $O^*(9^{\ecc+O(\log^2 \ecc)})$ time}
\DontPrintSemicolon
\label{alg:polyspace2}
\SetKwInOut{Input}{Input}\SetKwInOut{Output}{Output}
\Input{Connected graph $G$, an integer $\ecc$, and an integer $k$}
\Output{True if the treewidth of $G$ is at most $k$ and $G$ has an edge clique cover of size at most $\ecc$; true only if the treewidth of $G$ is at most $k$}
\lIf{$|\PC(G)| > 3^\ecc$}{
  \Return False
}
\For{$\pcl \in \PC(G)$}{
  \If{$|\pcl| \le k+1$}{
    \lIf{$|\CC(G \setminus \pcl)| > \ecc$}{
    \Return False
    }
    ok $\gets$ True\;
    \For{$C \in \CC(G \setminus \pcl)$}{
      \If{Treewidth$(R(C), \ecc/2+1, k)$ = False}{
        ok $\gets$ False\;
      }
    }
    \lIf{ok}{
      \Return True
    }
  }
}
\Return False\;
\end{algorithm}

The main idea to make the algorithm work when we do not know the edge clique cover is to just check if there are at most $3^\ecc$ PMCs.
The reason why the time complexity becomes a bit higher than in Lemma~\ref{lem:poly_alg1} is that we cannot assume that the worst case of branching from a PMC $\pcl$ results in only two subproblems.
In particular, we cannot assume anything better than $\ecc$ subproblems each with a minimum edge clique cover of size $\ecc/2+1$.

Algorithm~\ref{alg:polyspace2} presents a polynomial space $O^*(9^{\ecc + O(\log^2 \ecc)})$ time algorithm for treewidth.
The algorithm for weighted minimum fill-in is similar.

\begin{mylemma}
\label{lem:poly_alg2}
There is an algorithm that given a graph $G$ and an integer $\ecc$ uses polynomial space and $O^*(9^{\ecc + O(\log^2 \ecc)})$ time and returns an integer $t$ that is at least the treewidth of $G$.
If also a weight function $w : V(G)^2 \rightarrow \mathbb{R}_{\ge 0}$ is given, the algorithm also returns a number $f$ that is at least the weighted minimum fill-in of $G$ with respect to $w$.
If $\ecc$ is at least the size of a minimum edge clique cover of $G$, then $t$ is the treewidth of $G$ and $f$ is the weighted minimum fill-in of $G$ with respect to $w$.
\end{mylemma}
\begin{proof}
We use the same algorithm as in Lemma~\ref{lem:poly_alg1}, but with the modification that instead of giving an edge clique cover as a parameter we give just the integer $\ecc$.
At the start we check if $|\PC(G)| > 3^{\ecc}$ in $O^*(9^\ecc)$ time, and therefore we can use the same $3^{\ecc}$ bound on PMCs.
Also, each time we recurse from a PMC $\pcl$ we check that $|\CC(G \setminus \pcl)| \le \ecc$.
If not, then $G$ has no edge clique cover of size $\ecc$.

We use $k = \ecc$ and use induction on $k$.
Assume that the time complexity is $p(n) 9^{k+3 \log_2 k} k^{3 \log_2 k}$ with the same assumptions on $p(n)$ as in the proof of Lemma~\ref{lem:poly_alg1}.
We recurse from at most $3^k$ PMCs $\pcl$, from each into $|\CC(G \setminus \pcl)| \le k$ subproblems with the value of $k$ decreasing to $k/2+1$ in each.
Now, the time complexity is 
\begin{align*}
&p(n)9^k + 3^k k p(n) 9^{k/2+1 + 3\log_2(k/2+1)} (k/2 + 1)^{3 \log_2 (k/2 + 1)}\\
=& p(n) 9^k + p(n) 9^{k + 1 + 3 \log_2(k+2)-3} k(k/2+1)^{3 \log_2(k+2)-3}\\
\le& p(n) 9^k + p(n) 9^{k + 3 \log_2 k} k^{3 \log_2 k} / 9 \le p(n) 9^{k + 3 \log_2 k} k^{3 \log_2 k},
\end{align*}
which verifies the induction assumption.
\end{proof}

Theorem~\ref{the:poly_alg2} follows.

\section{Tightness}
\label{sec:tight}
We show that the bounds $S(\ecc, 2)$ and $S(\ecc, 3) + \ecc$ for the numbers of minimal separators and PMCs are tight for all values of $\ecc$.

Let us construct a graph $K^{\ecc}_2$.
Let $\w$ be a collection of size $\ecc$ initially containing disjoint sets of size 1, that is, $\w = \{W_1, \ldots, W_{\ecc}\}$ with $W_i = \{v_i\}$.
For each pair $1 \le i < j \le \ecc$ we insert an element $v_{i,j}$ into the sets $W_i$ and $W_j$.
Now, the vertex set of $K^{\ecc}_2$ is $V(K^{\ecc}_2) = \bigcup_{W \in \w} W$ and the edge set of $K^{\ecc}_2$ is $E(K^{\ecc}_2) = \bigcup_{W \in \w} W^2$.
The collection $\w$ is therefore an edge clique cover of $K^{\ecc}_2$.

\begin{mylemma}
The graph $K^{\ecc}_2$ has at least $S(\ecc, 2)$ minimal separators and at least $S(\ecc, 3) + \ecc$ PMCs.
\end{mylemma}
\begin{proof}
For distinct subsets $\w' \subseteq \w$ the vertex sets $V(K^{\ecc}_2, \w')$ are distinct because they can be identified by the inclusion of the $v_i$ vertices.
Let $\w'$ be any non-empty strict subset of $\w$.
Both $K^{\ecc}_2[V(K^{\ecc}_2, \w')]$ and $K^{\ecc}_2[V(K^{\ecc}_2, \w \setminus \w')]$ are connected graphs.
Any vertex in $V(K^{\ecc}_2) \setminus V(K^{\ecc}_2, \w', \w \setminus \w')$ is of type $v_{i,j}$ and has a neighbor in both $V(K^{\ecc}_2, \w')$ and $V(K^{\ecc}_2, \w \setminus \w')$.
Therefore, $V(K^{\ecc}_2) \setminus V(K^{\ecc}_2, \w', \w \setminus \w')$ is a minimal separator and $V(K^{\ecc}_2, \w')$ and $V(K^{\ecc}_2, \w \setminus \w')$ are blocks of it.
Therefore, the number of minimal separators of $K^{\ecc}_2$ is at least the number of bipartitions of $\w$.

Take any tripartition $\{\w_1, \w_2, \w_3\}$ of $\w$.
Note that distinct tripartitions define distinct sets $V(K^{\ecc}_2, \w_1, \w_2, \w_3)$.
We show that $\pcl = V(K^{\ecc}_2) \setminus (K^{\ecc}_2, \w_1, \w_2, \w_3)$ is a PMC of $K^{\ecc}_2$.
Any vertex in $\pcl$ is of type $v_{i,j}$, with vertices $v_i$ and $v_j$ in different blocks $V(K^{\ecc}_2, \w_p)$.
Therefore, for any pair of vertices in $\pcl$ there is a common block that they are adjacent to, and therefore $\pcl$ satisfies the cliquish condition.
A component $V(K^{\ecc}_2, \w_1)$ cannot be full because there is a vertex $v_{i,j}$ that intersects cliques $W_i \in \w_2$ and $W_j \in \w_3$ and therefore is in the PMC but not in the neighborhood of $V(K^{\ecc}_2, \w_1)$.
Therefore, $\pcl$ satisfies the no full component condition.
Therefore, the number of PMCs of $K^{\ecc}_2$ that contain no vertices of type $v_i$ is at least the number of tripartitions of $\w$.

We complete the proof by showing that for each $v_i$ the set $N[v_i]$ is a PMC, contributing the term $\ecc$ to the bound.
The set $N[v_i]$ satisfies the no full component condition because $v_i$ is not connected to any vertex outside of it.
It satisfies the cliquish condition because it is a clique.
\end{proof}

\section{Relation of Edge Clique Cover and Modular Width}
\label{sec:mw}
In~\cite{DBLP:journals/algorithmica/FominLMT18} an algorithm with time complexity $O^*(1.7347^\mw)$, where $\mw$ is the modular width, was given for enumerating PMCs.
In Section~\ref{sec:relwork} we claimed that the parameters edge clique cover and modular width have a relation $\mw \le 2^\ecc - 2$, which is tight.
Now we prove this claim.

A module of a graph $G$ is a vertex set $X \subseteq V(G)$ such that for all $x \in V(G) \setminus X$, either $N(x) \cap X = X$ or $N(x) \cap X = \emptyset$.
We use a recursive definition of modular width with four operations.

\begin{mydefinition}[\cite{DBLP:journals/algorithmica/FominLMT18}]
A graph $G$ has modular width $\mw$ if
\begin{enumerate}
\item $|V(G)| \le 1$, 
\item $G$ is a disjoint union of two graphs with modular width at most $\mw$,
\item $G$ is a join of two graphs $G_1$ and $G_2$ with modular width at most $\mw$, that is, $V(G) = V(G_1) \cup V(G_2)$ and $E(G) = E(G_1) \cup E(G_2) \cup (V(G_1) \times V(G_2))$, or
\item the vertices of $G$ can be partitioned into $k \le \mw$ modules $X_1, \ldots, X_k$, such that each $G[X_i]$ has modular width at most $\mw$.
\end{enumerate}
\end{mydefinition}

Now we prove the inequality $\mw \le 2^\ecc -2$.

\begin{mylemma}
If a graph has an edge clique cover of size $\ecc$ then it has modular width at most $2^\ecc - 2$.
\end{mylemma}
\begin{proof}
Let $G$ be a graph with edge clique cover $\w$ of size $\ecc$.
If $G$ has a vertex $v$ with $\w[v] = \emptyset$, then we remove $v$ with the operation~2.
If $G$ has a vertex $v$ with $\w[v] = \w$, then we remove $v$ with the operation~3.
Now vertices of $G$ can be partitioned to at most $2^\ecc-2$ classes based on $\w[v]$.
These classes are cliques and they are modules because if $\w[v] = \w[u]$ then $N[v] = N[u]$.
\end{proof}

Next we prove the tightness of this inequality.
\begin{mylemma}
For each integer $\ecc \ge 3$ there is a graph with an edge clique cover of size $\ecc$ and modular width $2^\ecc - 2$.
\end{mylemma}
\begin{proof}
We construct a graph $K^{\ecc}$.
Let $\w$ be a collection of size $\ecc$ that will be the edge clique cover of $K^{\ecc}$.
For every non-empty strict subset $\w' \subsetneq \w$ we create a vertex $v_{\w'}$ and insert it into each clique $W \in \w'$.
The edges of $K^{\ecc}$ are defined by $\w$, that is, by whether the subsets corresponding to the vertices intersect.

The graph $K^{\ecc}$ is connected and its complement is also connected, so operations~2 and~3 cannot be applied.
Next we prove that operation~4 cannot be applied with any integer $k < 2^\ecc-2$.
Suppose there is a non-trivial module $M$ of $K^{\ecc}$, i.e., a module with size $2 \le |M| < |V(K^{\ecc})|$.
Let $v_{\w_1}$ and $v_{\w_2}$ be two distinct vertices in $M$ with $|\w_1| \le |\w_2|$.
Now $v_{\w \setminus \w_1} \in M$, because $v_{\w \setminus \w_1} \in N(v_{\w_2})$, but $v_{\w \setminus \w_1} \notin N(v_{\w_1})$.
Now, since $M$ contains both $v_{\w_1}$ and $v_{\w \setminus \w_1}$ it must contain all vertices corresponding to single element subsets of $\w$ by a similar argument.
Since $M$ contains all vertices corresponding to single element subsets of $\w$ it also must contain all vertices of $K^{\ecc}$ by a similar argument.
\end{proof}

\section{Conclusion}
\label{sec:conclusion}
We bounded the number of minimal separators and PMCs by the size of a minimum edge clique cover, obtaining new parameterized algorithms for problems in the PMC framework.
The parameterization by edge clique cover is motivated by real applications of optimal triangulations, and our results provide theoretical corroboration on the observations of the efficiency of the PMC framework in practice.
Prior to our work, only the work of Fomin et al.~\cite{DBLP:journals/algorithmica/FominLMT18} considers FPT bounds for PMCs.
Our work answers to their proposal for finding further FPT parameterizations for PMCs.

We showed that our combinatorial bounds are the best possible, implying also that our PMC enumeration algorithm is optimal up to polynomial factors with respect to the parameter $\ecc$.
For individual problems it remains as an open problem to improve the algorithms or to prove conditional lower bounds assuming conjectures such as strong exponential time hypothesis~\cite{DBLP:journals/eatcs/LokshtanovMS11}.
We are not aware of other graph parameters whose value is always at most $\ecc$ and for which single-exponential FPT bounds for minimal separators and PMCs exist.
In particular, we note that graphs with vertex clique cover of size $2$ can have an exponential number of minimal separators: the graph consisting of two cliques of size $n/2$ connected by a matching of $n/2$ edges has $2^{n/2}-2$ minimal separators.

One combinatorial question closely related to our work is whether the bound $O(1.7347^n)$ for the number of PMCs can be improved in graphs where $\ecc \le n$.
This is motivated by moral graphs of Bayesian networks, for which the inequality holds.
Also the question of finding other useful parameterizations for PMCs still remains for future work.
Because of the fact that the parameter $\ecc$ occurs naturally in multiple settings related to optimal triangulations, we expect that even more applications of our results could arise in future.

\begin{acknowledgements}
I wish to thank Matti J\"arvisalo, Mikko Koivisto, Andreas Niskanen, and Juha Harviainen for helpful comments.
\end{acknowledgements}

%
\section*{Conflict of interest}
The authors declare that they have no conflict of interest.

\bibliographystyle{spmpsci}      
\bibliography{paper}   

\end{document}